\documentclass[12pt,nofootinbib,preprint,superscriptaddress]{revtex4}
\pdfoutput=1

\usepackage{amsmath}
\usepackage{amssymb}
\usepackage{anysize}
\usepackage{bold-extra}
\usepackage{color}
\usepackage{enumerate}
\usepackage{fancyhdr}
\usepackage{graphicx}
\usepackage[linktocpage,colorlinks,urlcolor=blue]{hyperref}
\usepackage{multirow}
\usepackage[final]{pdfpages}
\usepackage{rotating}
\usepackage{subfig}
\usepackage{textcomp}
\usepackage{url}
\usepackage{verbatim}
\usepackage{wrapfig}
\usepackage{amsfonts}
\usepackage{comment}
\usepackage{epigraph}
\usepackage[utf8]{inputenc}
\usepackage{slashed}
\usepackage{bbm}
\usepackage{cancel}
\usepackage{epsf, array, color}
\def\lsim{\mbox{\raisebox{-.6ex}{~$\stackrel{<}{\sim}$~}}}
\def\gsim{\mbox{\raisebox{-.6ex}{~$\stackrel{>}{\sim}$~}}}
\newcommand{\La}{\mathcal{L}}
\begin{document}

\title{An NLO QCD effective field theory analysis of \boldmath $W^+_{}W^-_{}$
  production at the LHC including fermionic operators
}

\author{Julien Baglio}
\email{julien.baglio@uni-tuebingen.de}
\affiliation{Institute for Theoretical Physics, University of
  T\"ubingen, Auf der Morgenstelle 14, 72076 T\"ubingen, Germany}

\author{Sally Dawson}
\email{dawson@bnl.gov}
\affiliation{Department of Physics, Brookhaven National Laboratory, Upton, N.Y., 11973~ U.S.A.}

\author{Ian M. Lewis}
\email{ian.lewis@ku.edu}
\affiliation{Department of Physics and Astronomy, University of Kansas, Lawrence, Kansas, 66045~ U.S.A.}

\begin{abstract}
We study the impact of anomalous gauge boson and fermion couplings on
the production of $W^+W^-$ pairs at the LHC. Helicity amplitudes are
presented separately to demonstrate the sources of new physics
contributions and the impact of QCD and electroweak corrections. The
QCD corrections have important effects on the fits to anomalous couplings, in
particular  when one $W$ boson is
longitudinally polarized and the other is transversely polarized.  In
effective field theory language, we demonstrate that the dimension-6
approximation to constraining new physics effects in $W^+W^-$ pair
production fails at $p_T\sim 500-1000$~GeV.
\end{abstract}

\maketitle

\section{Introduction}
The $SU(2)\times U(1)$ structure of the electroweak sector of the
Standard Model  completely determines the $W^+W^-V$ interactions
($V=\gamma, Z $).   The amplitudes for the production of $W^+W^-$
pairs involve subtle cancellations between contributions that grow
with energy, and individual Feynman diagrams violate perturbative
unitarity~\cite{Gaemers:1978hg,Duncan:1985vj,Hagiwara:1986vm}. In
models with new high scale physics,  the form of these interactions
can be changed, potentially spoiling the cancellations that impose
unitarity conservation, and so the
pair production of gauge bosons can be extremely sensitive to new
physics interactions, providing a stringent test of the Standard Model (SM).
Precision constraints on anomalous $3-$gauge boson couplings have been
found from $e^+e^-\rightarrow W^+W^-$ measurements at
LEP-II~\cite{Schael:2013ita}, and even stronger constraints have been
derived at the LHC from $W^+W^-$
production~\cite{Falkowski:2014tna,Butter:2016cvz,Berthier:2016tkq,deBlas:2016ojx,Zhang:2016zsp}. The
experimental analyses~\cite{Khachatryan:2015sga,Aad:2016wpd}, however,
assume that all the new physics is in the $3-$gauge boson
couplings. In principle, both the fermion couplings to $Z$ bosons and
$W$ bosons could be altered, changing the results of the fits to new
physics
contributions~\cite{Falkowski:2014tna,Berthier:2016tkq,Zhang:2016zsp}. Although
the $Z-$fermion couplings are highly constrained by LEP data, they can
still have numerically significant effects on the fit to $W^+W^-$ pair
production.

A consistent theoretical analysis requires the use of effective
Lagrangian techniques. The new physics is parameterized as an operator
expansion in inverse powers of the high scale, $\Lambda$, where the new physics
is assumed to occur,
\begin{equation}
\La_{±\rm SMEFT}=
\La\sim \La_{\rm SM}+\sum_{i,n}{C_i^{(n)}\over\Lambda^{n-4}}O_i^{(n)}+\ldots\,
\label{eq:smeft}
\end{equation}
where $O_i^{(n)}$ has mass dimension-$n$ and $\La_{\rm SM}$ contains
the complete SM Lagrangian. Neglecting flavor, there are 59
possible operators at
dimension-6~\cite{Buchmuller:1985jz,Grzadkowski:2010es}, but only a
small subset contribute to $W^+W^-$ production. The goal of this work
is to consistently extract limits on potential new physics effects in pair
production of $W^+W^-$ at the LHC, including modifications to
both three-gauge-boson vertices and fermion$-$gauge-boson vertices. We first
review the effects of non-SM interactions in the various helicity
channels, since the $W^+W^-$ helicity amplitudes have differing
behaviors at high energy, which may facilitate the extraction of
anomalous
couplings~\cite{Gaemers:1978hg,Duncan:1985vj,Hagiwara:1986vm,Azatov:2016sqh,Falkowski:2016cxu}.

The effects of new physics contributions to gauge boson pair
production can be expected to be of the same order of magnitude as QCD
and electroweak (EW) corrections, and so these contributions must be
included when extracting limits on new physics. The SM QCD corrections
to $W^+W^-$ pair production are known to
NNLO~\cite{Gehrmann:2014fva,Grazzini:2016ctr}, including the effects
of a jet veto~\cite{Dawson:2016ysj,Hamilton:2016bfu}. The EW
corrections are typically
small~\cite{Bierweiler:2013dja,Baglio:2013toa,Biedermann:2016guo}, and
the combined QCD/EW corrections including leptonic $W$ decays have
been
implemented~\cite{Biedermann:2016hmb,Biedermann:2016lvg,Kallweit:2017khh}. We perform
an analysis including QCD~\cite{Dixon:1999di} and EW
corrections~\cite{Baglio:2013toa}, along with modifications of both
the three-gauge-boson and fermion couplings. Section~\ref{sec:basics}
reviews the formalism of anomalous couplings in $W^+W^-$ pair
production and lowest order (LO) and next-to-leading order (NLO)
results are presented in Sections~\ref{sec:lo} and
\ref{sec:nlo}. Section~\ref{sec:conc} contains some conclusions about
the impact of our work on fits to anomalous couplings.
\section{Basics}
\label{sec:basics}
Assuming CP conservation, the most general Lorentz invariant $3-$gauge
boson couplings can be written
as~\cite{Gaemers:1978hg,Hagiwara:1986vm}
\begin{eqnarray}
 \La_{V}=
-ig_{WWV}\left(g_1^V\left(W^+_{\mu\nu}W^{-\mu}V^\nu-W_{\mu\nu}^-W^{+\mu}V^\nu\right)+\kappa^VW^+_\mu
            W^-_\nu V^{\mu\nu}+\frac{\lambda^V}{M^2_W}W^+_{\rho\mu}{W^{-\mu}}_\nu V^{\nu\rho}\right),
\label{eq:lagdef}
\end{eqnarray}  
where $V=\gamma, Z$ and $g_{WW\gamma}=e$ and $g_{WWZ}=g \cos\theta_W$,
$\theta_W^{}$ being the weak mixing angle. We use  the
abbreviations $s_W^{} \equiv \sin\theta_W^{}$ and $c_W^{} \equiv
\cos\theta_W^{}$. The fields in Eq.(\ref{eq:lagdef}) are the
canonically normalized mass eigenstate fields. In a similar fashion,
we define the effective couplings of fermions to gauge fields
and assume that the structure of the charged and neutral currents is that
of the SM\footnote{Dipole operators change the structure of the
  charged and neutral currents. However, the dipole contributions
  appear at dimension-8 in the amplitude squared since they do not
  interfere with the SM.  Hence, we neglect them.},
\begin{eqnarray}
  \La&=&g_ZZ_\mu\biggl[g_L^{Zq}+\delta g_{L}^{Zq}\biggr]
  {\overline q}_L\gamma_\mu q_L\
 +g_ZZ_\mu\biggl[g_R^{Zq}+\delta g_{R}^{Zq}\biggr]
  {\overline q}_R\gamma_\mu q_R\nonumber \\
  &&+{g\over \sqrt{2}}\biggl\{W_\mu\biggl[(1+\delta g_{L}^W){\overline q}_L\gamma_\mu q_L^\prime
  +\delta g_R^W
  {\overline q}_R\gamma_\mu q_R^\prime\biggr] +h.c.\biggr\}\, ,
  \label{eq:dgdef}
  \end{eqnarray}
 where $g_Z=e/(c_W^{}s_W^{})\equiv g/c_W$, $Q_q$ is the electric
 charge of the quarks, and $q$ denotes up-type or down-type quarks.
 The SM quark couplings are:
\begin{eqnarray}
g_R^{Zq}&=&-s_W^2 Q_q\quad{\rm and}\quad g_L^{Zq}=T_3^q -s_W^2 Q_q,
\end{eqnarray}
where $T_3^q=\pm \displaystyle \frac{1}{2}$. For the $3-$gauge boson
couplings we define $g_1^V = 1+\delta g_1^V$, $\kappa_{}^V=
1+\delta\kappa_{}^V$, and in the SM $\delta g_1^V = \delta\kappa_{}^V
= \lambda_{}^V = 0$. Because of gauge invariance we always have
$\delta g_1^\gamma = 0$. We assume $SU(2)$ invariance, which relates
the coefficients,
\begin{eqnarray}
\delta g_L^W&=&\delta g_L^{Zf}-\delta g_L^{Zf'},
\nonumber \\
\delta g_1^Z&=& \delta \kappa_{}^Z+{s_W^2\over c_W^2}\delta \kappa_{}^\gamma,
\nonumber \\
\lambda_{}^\gamma &=& \lambda_{}^Z,
\label{eq:su2rel}
\end{eqnarray}
where $f$ denotes up-type quarks and $f'$  down-type quarks. 

The helicity amplitudes for $q {\overline q}\rightarrow W^+W^-$ have
been derived in many
places~\cite{Gaemers:1978hg,Hagiwara:1986vm,Azatov:2016sqh,Falkowski:2016cxu},
and we summarize the results in Appendix~\ref{appendixA}. In the high
energy limit, $s\gg M_Z^2$, the SM amplitude has the behavior
${\cal A}(q {\overline {q}}\rightarrow W^+W^- )\sim
{\cal{O}}(1)$. In the presence of anomalous couplings,  the leading
contribution in the high energy limit comes from the longitudinal
gauge boson amplitudes, resulting from the interference of the SM
amplitudes with the  non-SM contribution. The amplitude ${{\cal
    A}}_{ss'\lambda\lambda'}$ for ${\overline {q}}_s
q_{s^\prime}\rightarrow W^+_\lambda W^{-}_{\lambda '}$, where
$s,s',\lambda,\lambda'$ label the respective particle helicities, has
the high energy limit,
\begin{eqnarray}
  {{\cal A}}_{+-00}&\rightarrow&
                        {g^2 s\over 2 M_W^2}\sin\theta\biggl\{
                        \delta\kappa_{}^Z\biggl( s_W^2 Q_q^{} -
                        T_3^{q}\biggr) - s_W^2 Q_q^{}
                        \delta\kappa_{}^\gamma - \delta g_{L}^{Zq} + 2
                        T_3^{q} \delta g_L^W\biggr\},\nonumber\\
{{\cal A}}_{-+00}&\rightarrow &
                       {g^2 s\over 2 M_W^2}\sin\theta\biggl\{
                       s_W^2 Q_q\biggl(\delta\kappa_{}^\gamma-\delta
                       \kappa_{}^Z\biggr) +\delta g_{R}^{Zq} \biggr\}\, .
 \label{eq:longlims}                    
\end{eqnarray}
We have retained only the linear contribution from the anomalous
couplings here and $\theta$ is the angle between the beam axis and the
gauge boson direction in the center-of-mass system. At high energies
the longitudinal amplitude coming from the non-SM couplings is
enhanced and is ${\cal O}(s/M_W^2)$, while the SM amplitude for
longitudinal $W^+W^-$ production is ${\cal{O}}(1)$. Hence, the
interference between SM and anomalous couplings is ${\cal O}(s/M_W^2)$
and grows with energy.

The SM and anomalous amplitudes for producing $2-$ transverse $W$
bosons with opposite helicities in the $\overline{q}_+q_-\rightarrow
W^+_{\pm} W^-_{\mp}$ configurations are ${\cal{O}}(1)$, while the SM
amplitudes with same helicity $W$ bosons in the
$\overline{q}_+q_{-}\rightarrow W^+_{\pm} W^-_{\pm}$ configurations
are ${\cal{O}}(M_W^2/s)$ and the leading term from the anomalous couplings is,
\begin{align}
  {{\cal A}}_{+-\pm\pm} \rightarrow -g^2 \lambda_{}^Z  T_3^q{s\over 2
  M_W^2}\sin\theta,
\label{eq:transgrow}
\end{align}
leading to a growth at high energies in the transverse amplitude in
the presence of non-zero $\lambda_{}^Z$.  The configurations with
right-handed quarks with same helicity $W$ bosons, $\overline{q}_- q_+
\rightarrow W^+_{\pm}W^-_{\pm}$, are $\mathcal{O}(1)$ and opposite
helicity $W$s, $\overline{q}_- q_+ \rightarrow W^+_{\pm}W^-_{\mp}$,
are zero.  Following this discussion, even though the anomalous
coupling amplitude grows with energy, it is clear that the
interference between SM and anomalous couplings is at most ${\cal
  O}(1)$.

Finally, the SM amplitude for producing one longitudinal and one
transverse gauge boson is suppressed by $M_W/\sqrt{s}$, while
the contribution from anomalous couplings is ${\cal O}(\sqrt{s}/M_W)$,
making this channel also quite sensitive to anomalous
couplings,
\begin{eqnarray}
 {\cal A}_{+-0\mp}={{\cal A}}_{+-\pm0}  & \rightarrow &  
                                    {g^2 \sqrt{s}\over {\sqrt{2}} M_W}
                                    (1\pm\cos\theta)
                                    \biggr\{
                                    \delta g^{Zq}_L - 2
                                    T_3^q \delta g^W_L +
                                    \frac{T_3^q}{2}\left(2\delta g_1^Z +
                                    \lambda_{}^Z -
                                    \frac{s_W^2}{c_W^2}\delta\kappa_{}^\gamma\right)
                                    +\nonumber\\
                    & & \phantom{ {g^2 \sqrt{s}\over {\sqrt{2}} M_W}
                        (1+\cos\theta) \biggr\{}
                        \frac{s_W^2
                        Q_q^{}}{2}\left(\frac{\delta\kappa_{}^\gamma}{c_W^2}-2\delta
                        g_1^Z\right)\biggr\},
 \nonumber \\
 {\cal A}_{-+0\mp}={{\cal A}}_{-+\pm0}  & \rightarrow &  
                                    \frac{g^2_{} \sqrt{s}}{\sqrt{2} M_W^{}}
                                    (1\mp\cos\theta)
                                    \biggr\{ \delta g_R^{Zq} +
                                    \frac{s_W^2 Q_q^{}}{2}
                                    \left({\delta\kappa_{}^\gamma
                                    \over c_W^2} - 2\delta
                                    g_1^Z\right)\biggr\} \, .
\label{eq:longtransgrow}
\end{eqnarray}
As with the the two transverse $W$ case, when one $W$ is transverse
and the other longitudinal the interference between SM and anomalous
couplings is $\mathcal{O}(1)$.

The Lagrangians of Eqs.(\ref{eq:lagdef}) and (\ref{eq:dgdef}) can be
mapped onto the  effective Lagrangian (EFT)  of Eq.(\ref{eq:smeft}),
where we work to dimension-$6$, assuming that the scale $\Lambda$ is
much larger than the weak scale, and that the couplings $C_i^{}$ are
perturbative. For simplicity, we work in the Warsaw
basis~\cite{Grzadkowski:2010es} and the dimension-$6$ operators
relevant for our analysis are,
\begin{eqnarray}
\mathcal{O}_{3W}&=& \epsilon^{abc} W_\mu^{a\nu}W_\nu^{b\rho}W_\rho^{c\mu},\nonumber \\
\mathcal{O}_{HD}&=& \mid \Phi^\dagger (D_\mu \Phi)\mid^2,\nonumber \\
\mathcal{O}_{HWB}&=& \Phi^\dagger\sigma^a\Phi W^a_{\mu\nu}B^{\mu\nu}\nonumber \\
\mathcal{O}_{HF}^{(3)}&=&i\biggl(\Phi ^\dagger \overleftrightarrow {D}_\mu^a \Phi \biggr) {\overline f}_L\gamma^\mu \sigma^a f_L,\nonumber\\ 
\mathcal{O}_{HF}^{(1)}&=&i\biggl(\Phi ^\dagger  \overleftrightarrow {D}_\mu \Phi \biggr){\overline f}_L \gamma^\mu  f_L,\nonumber \\
\mathcal{O}_{Hf}&=&i\biggl(\Phi ^\dagger  \overleftrightarrow{D}_\mu \Phi \biggr){\overline q}_R\gamma^\mu  q_R,\nonumber \\
\mathcal{O}_{Hud}&=&i\biggl(\widetilde{\Phi}^\dagger D_\mu \Phi\biggr) \overline{u}_R\gamma^\mu d_R,\nonumber\\
\mathcal{O}_{ll}&=&({\overline l}_L\gamma^\mu l_L)({\overline l}_L\gamma_\mu l_L),
\label{eq:ops}
\end{eqnarray}
where  $D_\mu \Phi=(\partial_\mu -i\,\frac{g}{2}\sigma^a
W^a_\mu-i\frac{g'}{2}B_\mu)\Phi$, $W^a_{\mu\nu}=\partial_\mu W^a_\nu
-\partial_\nu W^a_\mu+g\varepsilon^{abc}W^b_\mu W^c_\nu$,
$\Phi^\dagger \overleftrightarrow{D}_\mu \Phi=\Phi^\dagger D_\mu
\Phi-(D_\mu \Phi^\dagger)\Phi$, and $\Phi^\dagger
\overleftrightarrow{D}^a_\mu \Phi=\Phi^\dagger D_\mu
\sigma^a\Phi-(D_\mu \Phi^\dagger)\sigma^a\Phi$. $\Phi$ stands for the
Higgs doublet field with a vacuum expectation value $\langle\Phi\rangle =
(0,v/\sqrt{2})^{\rm T}$. The Lagrangian of Eq.(\ref{eq:smeft}) introduces non-canonically
normalized gauge fields. The input parameters we choose for our
analysis are $G_F^{}=1.16637\times 10^{-5}_{}$~GeV$^{-2}_{}$, $M_Z^{}=91.1876$~GeV and
$M_W^{}=80.385$~GeV, taken from their experimental values. In the
mapping from EFT operators to anomalous couplings we have to take into
account the EFT shifts $g_Z^{} \to g_Z^{} + \delta g_Z^{}$, $v\to v(1
+ \delta v)$, $s_W^{2} \to s_W^2 + \delta s_W^2$ in the definition of
the model input parameters for the gauge couplings, as well as for
$s_W^{}$, so that we get back to canonically normalized gauge
fields. This gives~\cite{Falkowski:2015fla,Brivio:2017bnu}
\begin{eqnarray}
\delta v &=& C_{Hl}^{(3)} - \frac12 C_{ll}^{},\nonumber\\
\delta g_Z^{} &=& -\frac{v^2}{\Lambda^2}\left(\delta v +\frac14 C_{HD}^{}\right),\nonumber\\
\delta s_W^2 &=& -\frac{v^2}{\Lambda^2} \frac{s_W^{} c_W^{}}{c_W^2-s_W^2}\left[2 s_W^{} c_W^{}\left(\delta v + \frac14 C_{HD}^{}\right) + C_{HWB}^{}\right],
\end{eqnarray}
where the tree-level relations are still valid:
\begin{eqnarray}
v^2=\frac{1}{\sqrt{2}G_F},\quad s^2_W=1-\frac{M_W^2}{M_Z^2},\quad g_Z=\frac{2M_Z}{v}=\frac{g}{c_W}=\frac{e}{c_W s_W}.
\end{eqnarray}
Using these shifts and the operators defined in Eq.(\ref{eq:ops}) we
find  the following mapping,
\begin{eqnarray}
\delta g_1^Z &=& \frac{v^2}{\Lambda^2}\frac{1}{c_W^2-s_W^2}\left(\frac{s_W^{}}{c_W^{}}C_{HWB}^{} + \frac14 C_{HD}^{} +\delta v\right),\nonumber\\
\delta \kappa_{}^Z &=& \frac{v^2}{\Lambda^2}\frac{1}{c_W^2-s_W^2}\left(2 s_W c_W C_{HWB}^{} + \frac14 C_{HD}^{} +\delta v\right),\nonumber\\
\delta \kappa_{}^\gamma &=& -\frac{v^2}{\Lambda^2}\frac{c_W^{}}{s_W^{}}C_{HWB}^{},\nonumber\\
\lambda_{}^\gamma &=& \frac{v}{\Lambda^2} 3 M_W^{} C_{3W},\nonumber\\
\lambda_{}^Z &=& \frac{v}{\Lambda^2} 3 M_W^{} C_{3W},\nonumber\\
\delta g_L^W &=& \frac{v^2}{\Lambda^2}C_{Hq}^{(3)} + c_W^2\delta g_Z^{} + \delta s_W^2,\nonumber\\ 
\delta g_R^W&=&\frac{v^2}{2\Lambda^2}C_{Hud}\nonumber\\
\delta
  g^{Zu}_L&=&-\frac{v^2}{2\Lambda^2}\left(C_{Hq}^{(1)}-C_{Hq}^{(3)}\right) + \frac12 \delta g_Z + \frac23\left(\delta s_W^2 - s_W^2 \delta g_Z^{}\right),\nonumber\\
\delta
  g^{Zd}_L&=&-\frac{v^2}{2\Lambda^2}\left(C_{Hq}^{(1)}+C_{Hq}^{(3)}\right) - \frac12 \delta g_Z - \frac13\left(\delta s_W^2 - s_W^2 \delta g_Z^{}\right),\nonumber\\
\delta g^{Zu}_R&=&-\frac{v^2}{2\Lambda^2} C_{Hu} + \frac23\left(\delta s_W^2 - s_W^2\delta g_Z^{}\right),\nonumber\\
\delta g^{Zd}_R&=&-\frac{v^2}{2\Lambda^2} C_{Hd} - \frac13\left(\delta s_W^2 - s_W^2\delta g_Z^{}\right),
\label{eq: mapcoef}
\end{eqnarray}
in agreement with Refs.~\cite{Berthier:2015oma,Zhang:2016zsp}

The operator $\mathcal{O}_{Hud}$ can mediate sources of flavor
violation in addition to the SM.  To suppress this new source, we work
under the assumption of minimal flavor violation
(MFV)~\cite{Chivukula:1987py,DAmbrosio:2002vsn}: all flavor violation
is generated via the Yukawa matrices. Under this assumption
$C_{Hud}\sim Y_u Y_d$, where $Y_u$ and $Y_d$ are up- and down-quark
Yukawa matrices, respectively~\cite{Alonso:2013hga}.  Hence, for light
initial state quarks we can safely assume $\delta
g_R^W=0$. Additionally, $C_{Hq}^{(3)},\, C_{Hq}^{(1)},$ and $C_{Hf}$
are assumed to be flavor diagonal and universal.

The amplitude  for $W^+W^-$ production is generically written as,
\begin{equation}
{{\cal A}}
\sim {{\cal A}}_{\rm SM}+{1\over \Lambda^2}{{\cal A}}_{\rm EFT}+\ldots
\end{equation}
In a consistent EFT approach, one should keep only the contributions
to the cross section proportional to $1/ \Lambda^2$, and so the
amplitude-squared is,
\begin{equation}
\sigma\sim {1\over s}\biggl(\mid {{\cal A}}_{\rm SM}\mid^2 + {{\cal A}}_{\rm SM}^* {
{{\cal A}}_{\rm EFT}\over \Lambda^2}+\ldots\biggr)
\end{equation}
Dropping the $1/\Lambda^4$ terms (and beyond) means that the cross
section is not guaranteed to be positive and the region of validity of
the EFT is hence restricted~\cite{Biekotter:2016ecg}. We will discuss
this in detail in the next section.

\section{Numerical Results}
\subsection{Lowest order and NLO electroweak effects}
\label{sec:lo}
\begin{figure}
  \centering
\subfloat{\includegraphics[width=0.48\textwidth]{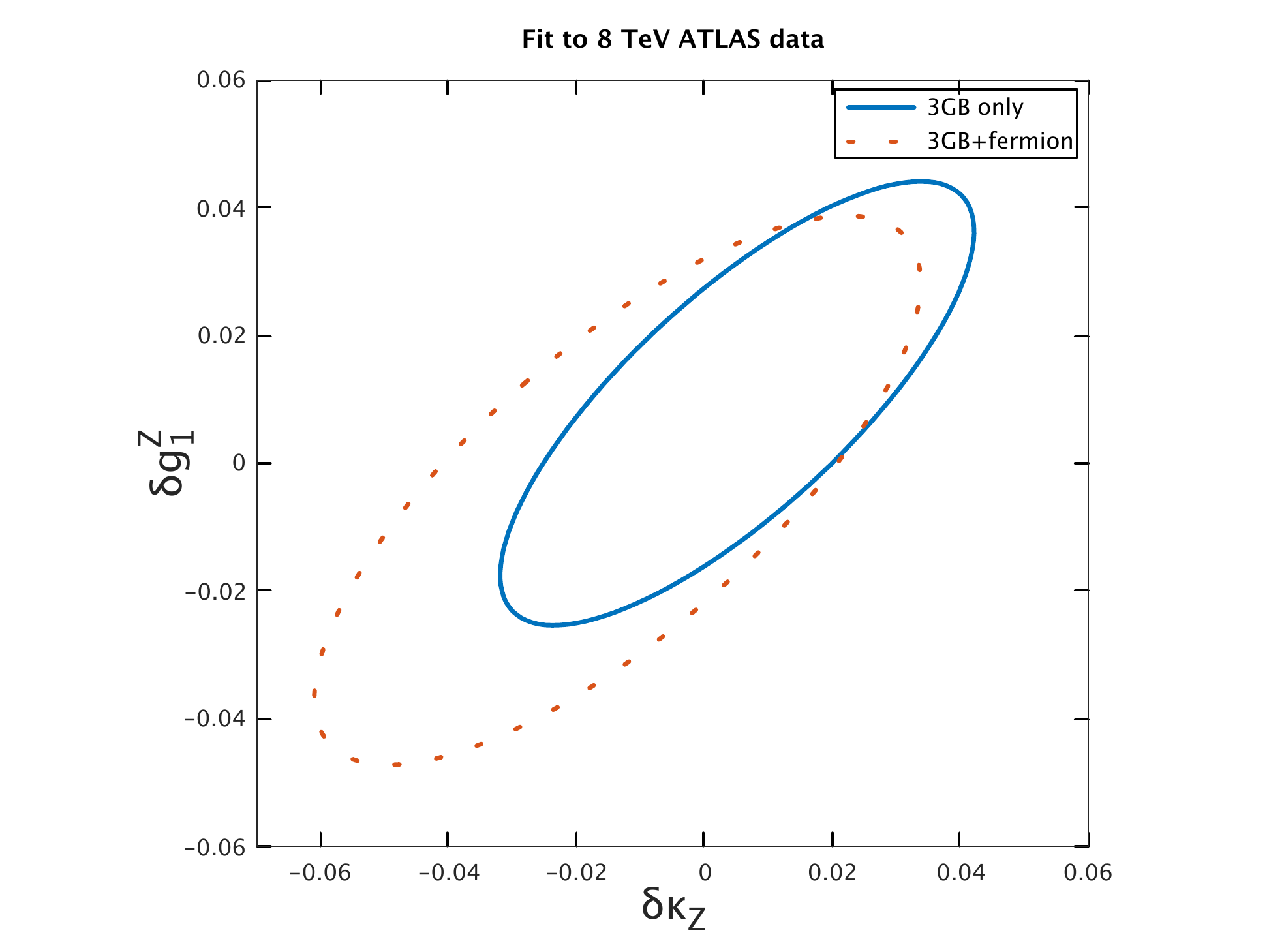}}
\subfloat{\includegraphics[width=0.48\textwidth]{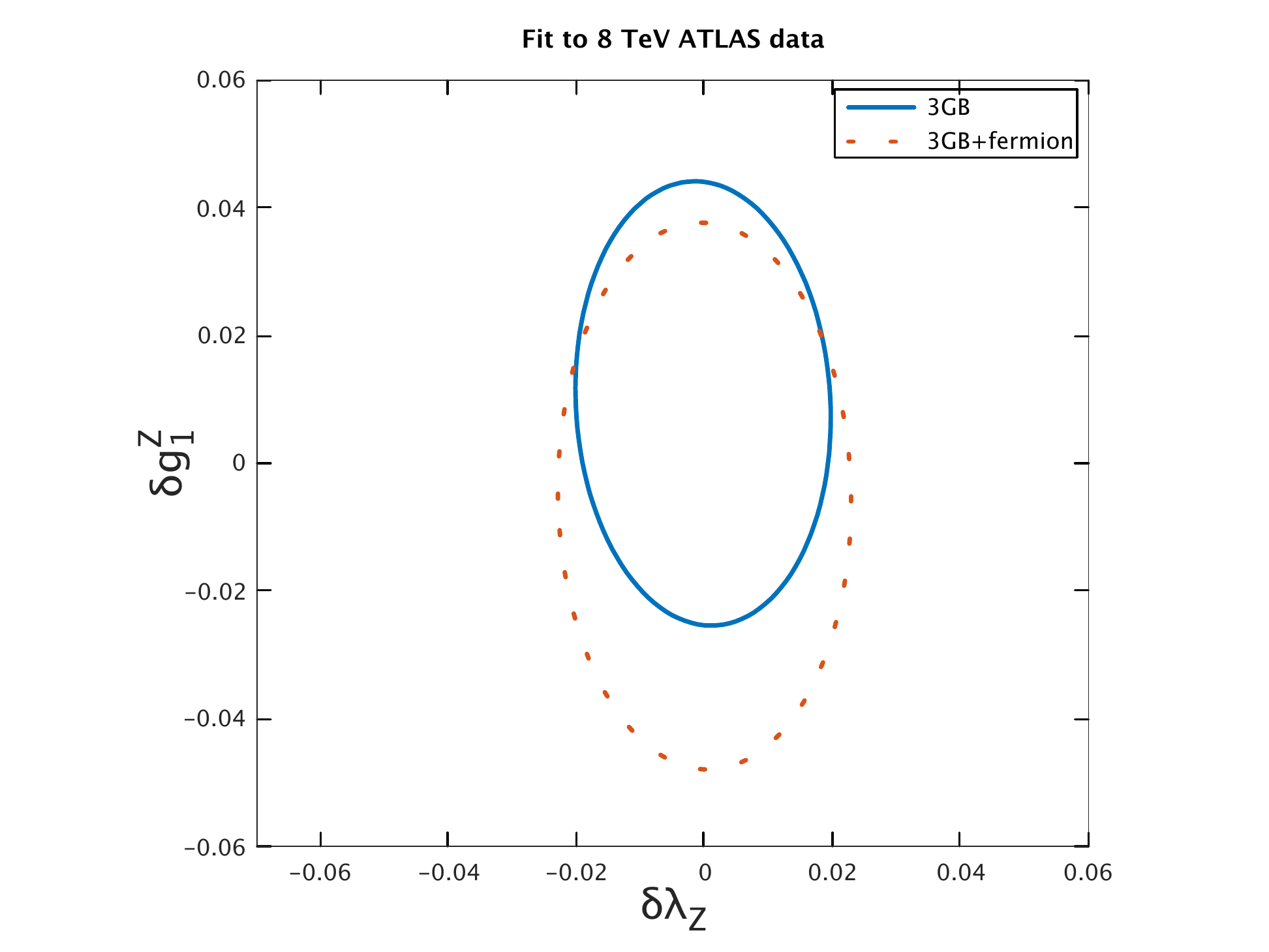}}\\
\subfloat{\includegraphics[width=0.48\textwidth]{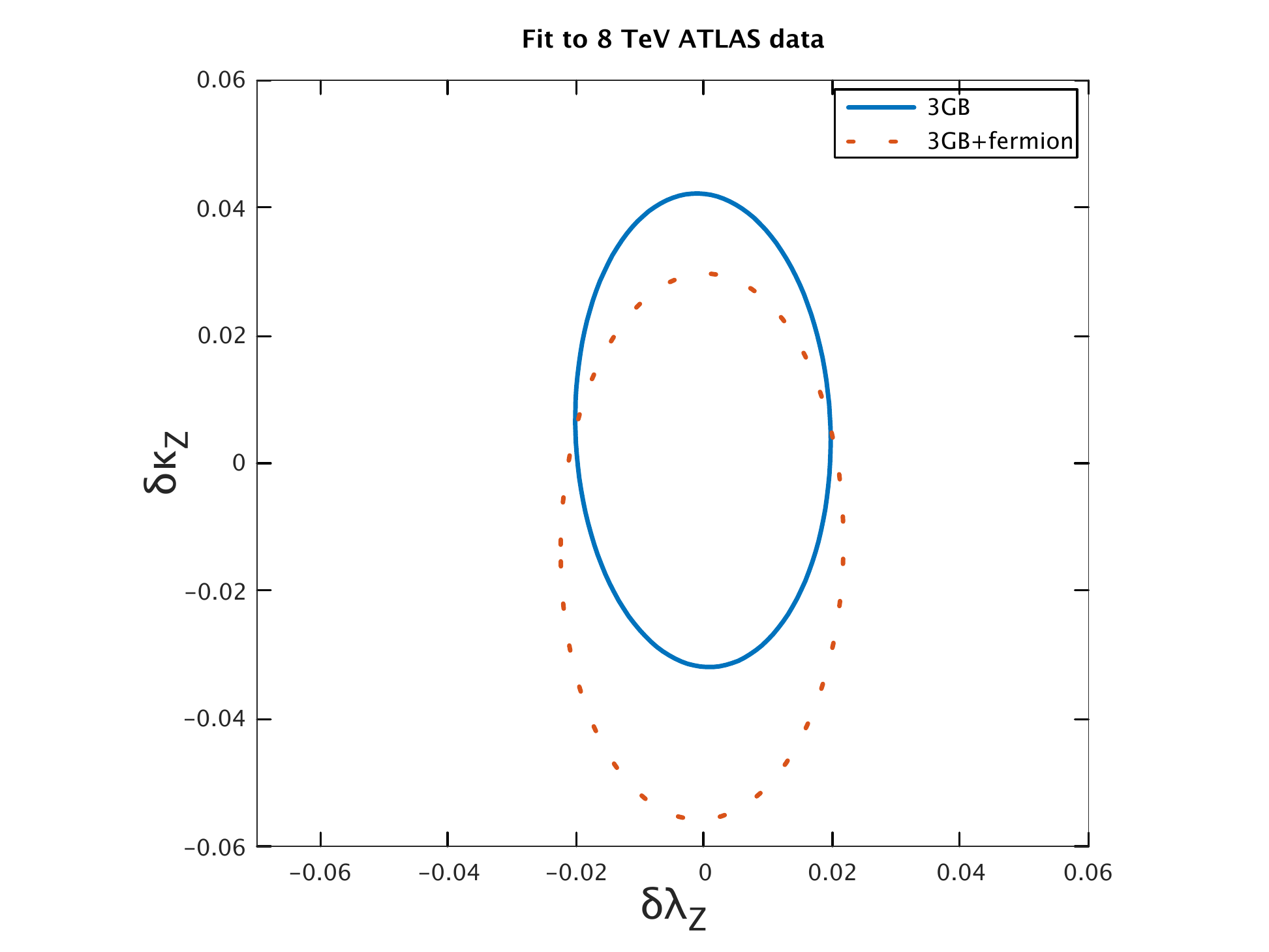}}
 \caption{Results of a parameter scan over (red dashed lines)
   anomalous triple gauge boson and fermion couplings and (solid
   blue lines) anomalous triple gauge boson couplings with fermion
   couplings set to their SM values.  The allowed regions are inside
   the ellipses.}
   \label{fig:atfig}
\end{figure}
It is well known that both anomalous fermion couplings and non-SM
three gauge boson couplings lead to cross sections which grow at high
energy, Eq.(\ref{eq:longlims}).  By fitting to the deviation of the
high $p_T$ spectrum from
the SM prediction, limits are obtained on the size of the anomalous
couplings.  Both ATLAS and CMS~\cite{Sirunyan:2017bey} have searched
for anomalous triple gauge boson couplings at the 8 TeV LHC. Using
the ATLAS bounds on the anomalous triple gauge boson couplings, at leading
order we determine a range of the cross section
\begin{eqnarray}
\sigma_{\rm min}^{\rm cut}<\sigma^{\rm cut}\equiv\sigma(p_T^{W^+}>500~{\rm GeV})=\int^\infty_{500~{\rm GeV}} dp_T^{W^+} \frac{d\sigma}{dp_T^{W^+}}<\sigma_{\rm max}^{\rm cut}~\label{eq:cut},
\end{eqnarray}
where $\sigma=\sigma(pp\rightarrow W^+ W^-)$ is the $W^+W^-$ hadronic
production cross section and $p_T^{W^+}$ is the $W^+$ transverse
momentum\footnote{We use the complete $\mid A\mid^2$ for the scans.}.  It is assumed that any $\sigma^{\rm cut}$ above
$\sigma_{\rm max}^{\rm cut}$ or below $\sigma_{\rm min}^{\rm cut}$
would have been observable and any point violating these bounds is
rejected. Using this technique, we reproduce both the ATLAS and CMS
bounds on anomalous $3-$gauge boson couplings. Next, we perform a scan
over all anomalous triple gauge boson couplings and fermion
couplings. The anomalous fermion couplings in our scan are constrained
by LEP limits~\cite{Falkowski:2014tna}:
\begin{eqnarray}
\delta g_L^{Zu}&=&(-2.6\pm 1.6)\times 10^{-3},\nonumber\\
\delta g_L^{Zd}&=&(2.3\pm 1)\times 10^{-3},\nonumber\\
\delta g_R^{Zu}&=& (-3.6\pm 3.5)\times 10^{-3},\nonumber\\
\delta g_R^{Zd}&=& (16.0\pm 5.2)\times 10^{-3}.
\end{eqnarray}
The results of these scans are shown in Fig.~\ref{fig:atfig}; the 
allowed regions are within the ellipses\footnote{The ellipses were
  determined using the Khachiyan Algorithm as implemented in
  Ref. \cite{ellipses}.}. The blue lines consider only anomalous
triple gauge boson couplings with fermion couplings set to the SM
values. These results are consistent with the ATLAS results. The red
lines consider both non-zero anomalous triple gauge boson and anomalous
fermion couplings. As can be seen, by including the anomalous fermion
couplings the central values of the allowed parameters change and the
areas of the allowed regions increase\footnote{Fits to CMS $W^+W^-$
  data lead to similar conclusions.}.  Although LEP constrains them to
be very small, the importance of the anomalous fermion couplings is
already apparent.  We have checked that these results are stable
against changes in the $p_T^{W^+}$ lower bound in Eq.(\ref{eq:cut}).

We consider two representative scenarios, allowed by global fits to
anomalous fermion couplings and anomalous 3-gauge-boson
couplings~\cite{Berthier:2015gja,Berthier:2016tkq,Falkowski:2014tna}
both at the same time. 
 The scenarios we consider are:
\begin{eqnarray}
  {\rm 3GB}:&\qquad \qquad &\delta g_1^Z =0.0163,\,
                             \delta\kappa_{}^Z=0.0239,\,
                             \lambda_{}^Z=0.00452,\nonumber\\
  {\rm Ferm}:&\qquad \quad & \delta g_{L}^{Zu}=-0.00239,\,
                             \delta g_{R}^{Zu}=-0.0069,\,\nonumber\\
            &\qquad \quad &  \delta
                            g_{L}^{Zd}=0.00271,\, \delta
                            g_{R}^{Zd}=0.0212. 
\label{eq:fitparm}
\end{eqnarray}
In addition, $\delta g^W_L$, $\lambda_{}^\gamma$, and $\delta
\kappa_{}^\gamma$ are determined by the relations from
Eq.(\ref{eq:su2rel}), and $\delta g^{W}_R$ is set to zero according to
our MFV assumption. In the ``3GB'' scenario we set the fermionic
anomalous couplings to zero (only the three-gauge-boson anomalous
couplings are considered), while in the ``Ferm'' scenario we set the
three-gauge-boson anomalous couplings to zero (only the fermionic
anomalous couplings are considered). The anomalous couplings of
Eq.(\ref{eq:fitparm}) can be translated to the EFT Wilson coefficients
using Eq.(\ref{eq: mapcoef}) for the two scenarios considered. In the
``Ferm'' scenario we have in general $C_{HWB}^{}=C_{3W}^{}=0$ as well
as $C_{HD}^{} = 2 C_{ll}^{} - 4 C_{Hl}^{(3)}$, while for the ``3GB''
scenario all coefficients are in principle non-zero in our operator
basis and have the following relations,
\begin{eqnarray}
  C_{Hu}^{} &=& 4 C_{Hq}^{(1)},\, C_{Hd}^{} = -2
  C_{Hq}^{(1)},\nonumber\\
  C_{Hq}^{(3)} &=&
  \frac{c_W^{}s_W^{}}{c_W^2-s_W^2}\left\{C_{HWB}^{}+{c_W^{} \over
                   s_W^{}}\left(\frac14 C_{HD}^{}+C_{Hl}^{(3)}-\frac12
                   C_{ll}^{}\right)\right\}.
\end{eqnarray}
We obtain the following Wilson coefficients in the ``3GB'' scenario,
\begin{eqnarray}
  \frac{1}{\Lambda^2_{}}\left(C_{Hl}^{(3)} -\frac12 C_{ll}^{} +
  \frac14 C_{HD}^{}\right) = 2.36 \times
  10^{-8}_{}~\text{GeV}^{-2}_{},\nonumber\\
  {C_{3W}^{}\over \Lambda^2_{}} = 7.61\times
  10^{-8}_{}~\text{GeV}^{-2}_{},\,
  {C_{HWB}^{} \over \Lambda^2_{}} = 2.34\times
  10^{-7}_{}~\text{GeV}^{-2}_{},\nonumber\\
  {C_{Hq}^{(1)}\over \Lambda^2_{}} = -6.18\times
  10^{-8}_{}~\text{GeV}^{-2}_{},\, 
  {C_{Hq}^{(3)}\over \Lambda^2_{}} =  2.09\times
  10^{-7}_{}~\text{GeV}^{-2}_{},\nonumber\\
  {C_{Hu}^{}\over \Lambda^2_{}} = -2.47\times
  10^{-7}_{}~\text{GeV}^{-2}_{},\,
  {C_{Hd}^{}\over \Lambda^2_{}} =  1.24\times
  10^{-7}_{}~\text{GeV}^{-2}_{},
\label{eq:eftparms3GB}
\end{eqnarray}
and we obtain the following Wilson coefficients in the ``Ferm''
scenario,
\begin{eqnarray}
  \frac{1}{\Lambda^2_{}}\left(C_{Hl}^{(3)} -\frac12 C_{ll}^{} +
  \frac14 C_{HD}^{}\right) &=& 0,\nonumber\\
  {C_{3W}^{}\over \Lambda^2_{}} = {C_{HWB}^{} \over \Lambda^2_{}} &=&
  0,\nonumber\\
  {C_{Hq}^{(1)}\over \Lambda^2_{}} = -5.28\times
  10^{-9}_{}~\text{GeV}^{-2}_{}&,\,& 
  {C_{Hq}^{(3)}\over \Lambda^2_{}} =  -8.41\times
  10^{-8}_{}~\text{GeV}^{-2}_{},\nonumber\\
  {C_{Hu}^{}\over \Lambda^2_{}} = 2.28\times
  10^{-7}_{}~\text{GeV}^{-2}_{}&,\,&
  {C_{Hd}^{}\over \Lambda^2_{}} =  -6.99\times
  10^{-7}_{}~\text{GeV}^{-2}_{}.
\label{eq:eftparmsFerm}
\end{eqnarray}
Assuming that all $C_i^{}$ are perturbative ($C_i\lsim
1$), the lower bound on the EFT scale is $\Lambda \gsim 2.8$~TeV. 

\begin{figure}
  \centering
\subfloat{\includegraphics[width=0.48\textwidth]{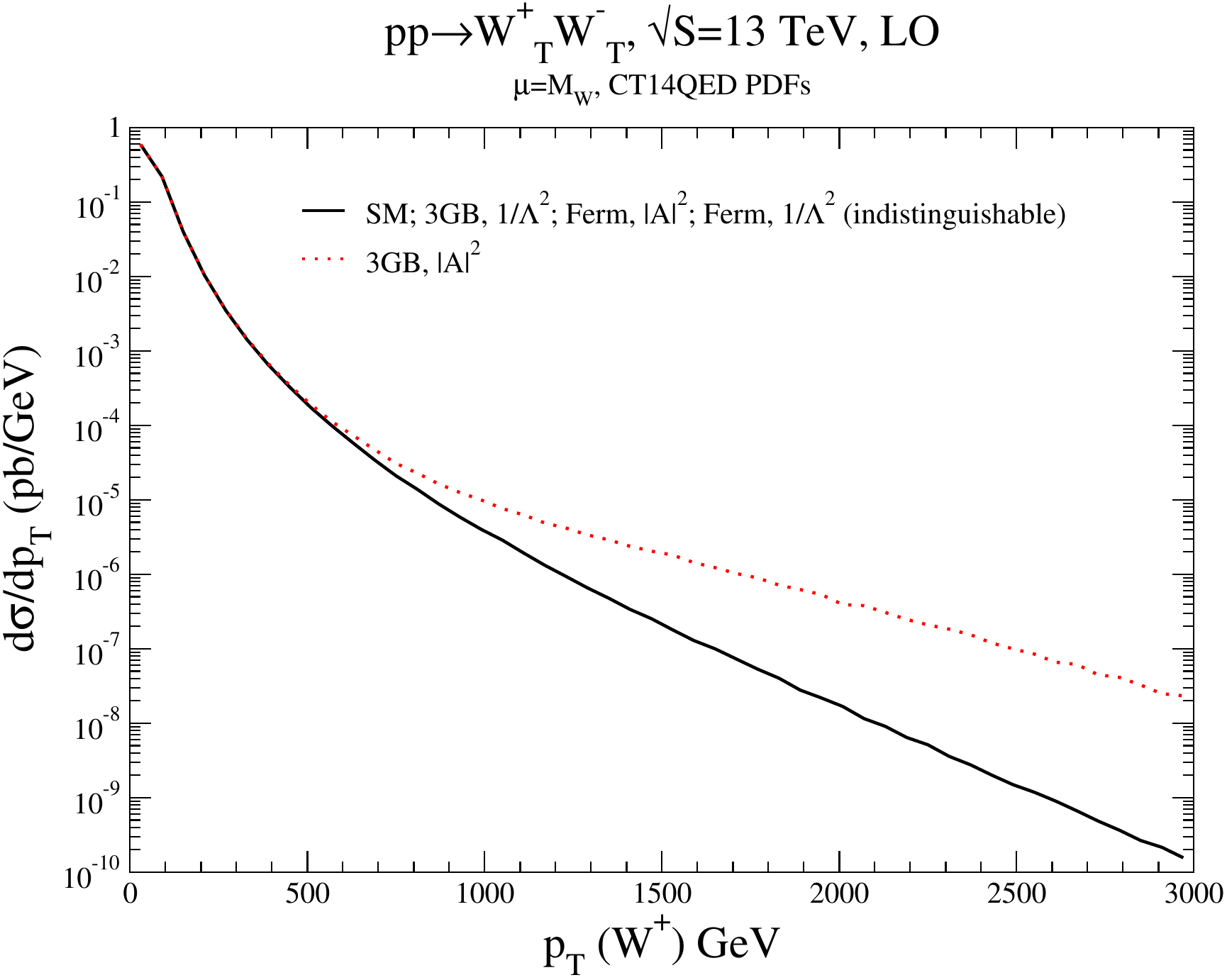}}
\subfloat{\includegraphics[width=0.48\textwidth]{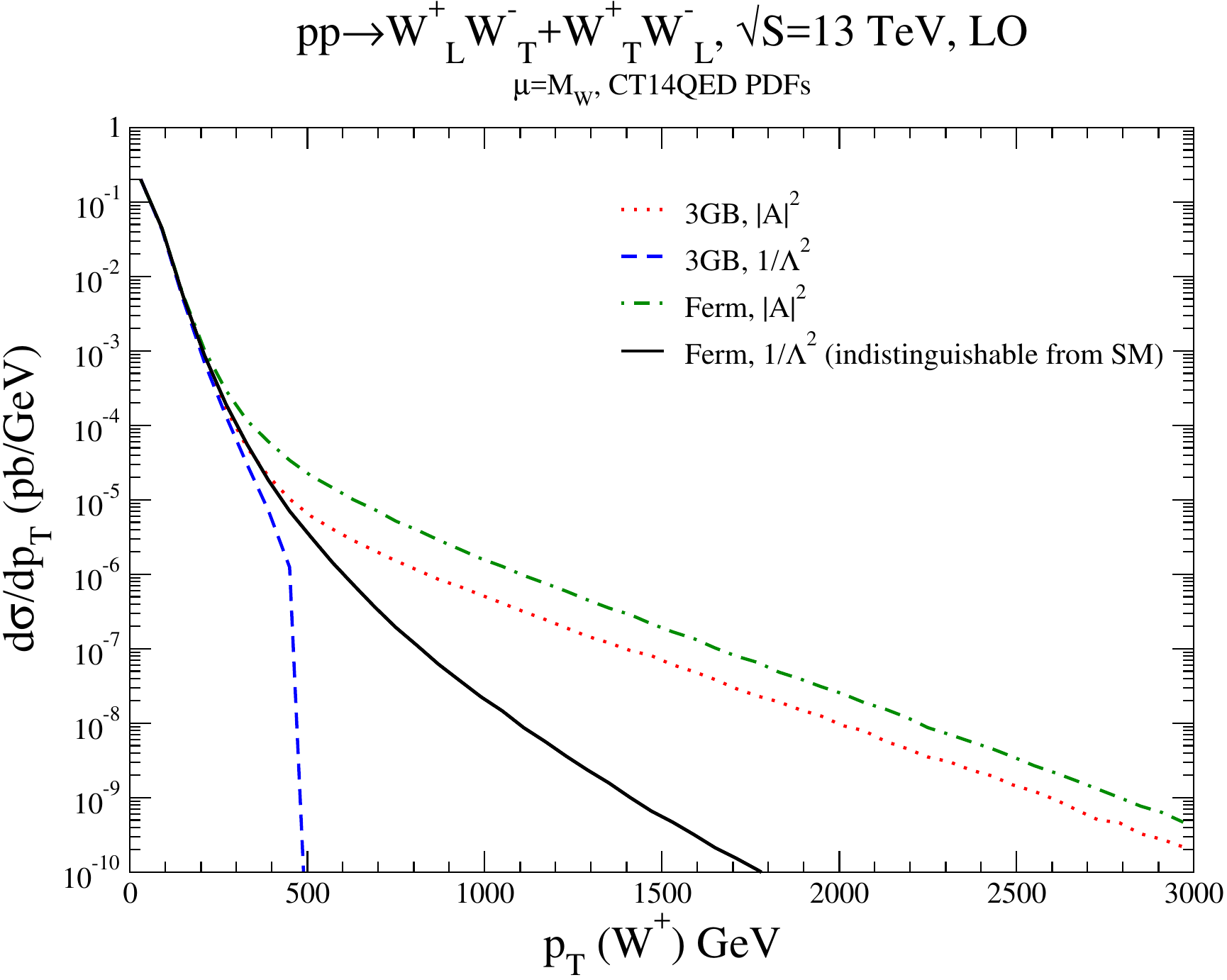}}
 \caption{Tree-level cross sections for $W^+_TW^-_T$ (LHS) and
   $W^+_TW^-_L +W^+_LW^-_T$ (RHS) production at the 13 TeV LHC in the
   SM and in the scenarios of Eq.(\ref{eq:fitparm}). The curves
   labelled $|A|^2$ include the square of the dimension-6
   amplitudes, while the curves labelled $1/\Lambda^2$ have the EFT
   result consistently truncated at  $1/\Lambda^2$.  The LT curve
   (RHS) for the 3GB amplitude in the EFT should be truncated at
   $p_T\sim 500$~GeV, where the  LT rate becomes negative and the EFT
   expansion fails.}
   \label{fig:ptamps}
\end{figure}
\begin{figure}
  \centering
\subfloat{\includegraphics[width=0.48\textwidth]{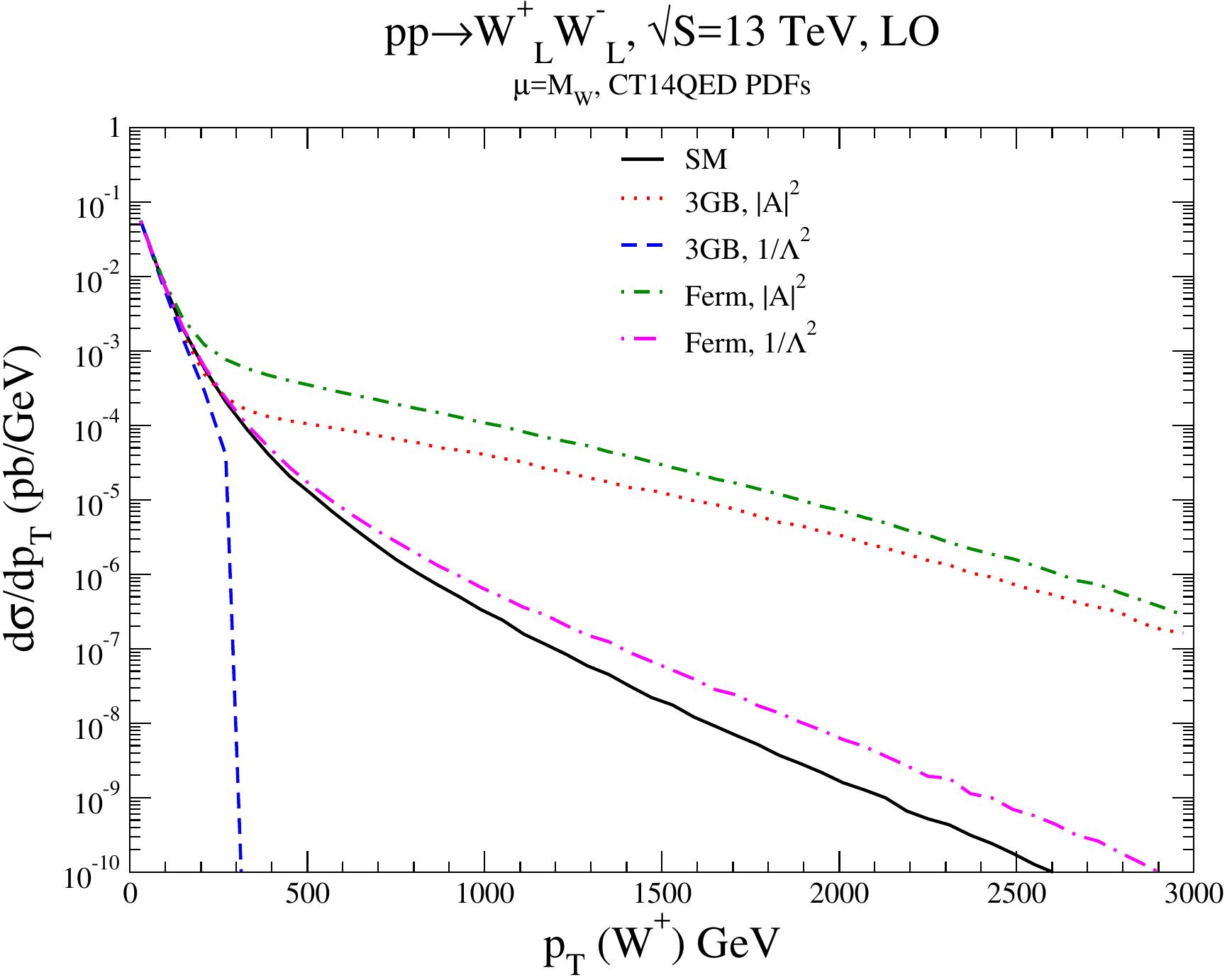}}
\subfloat{\includegraphics[width=0.48\textwidth]{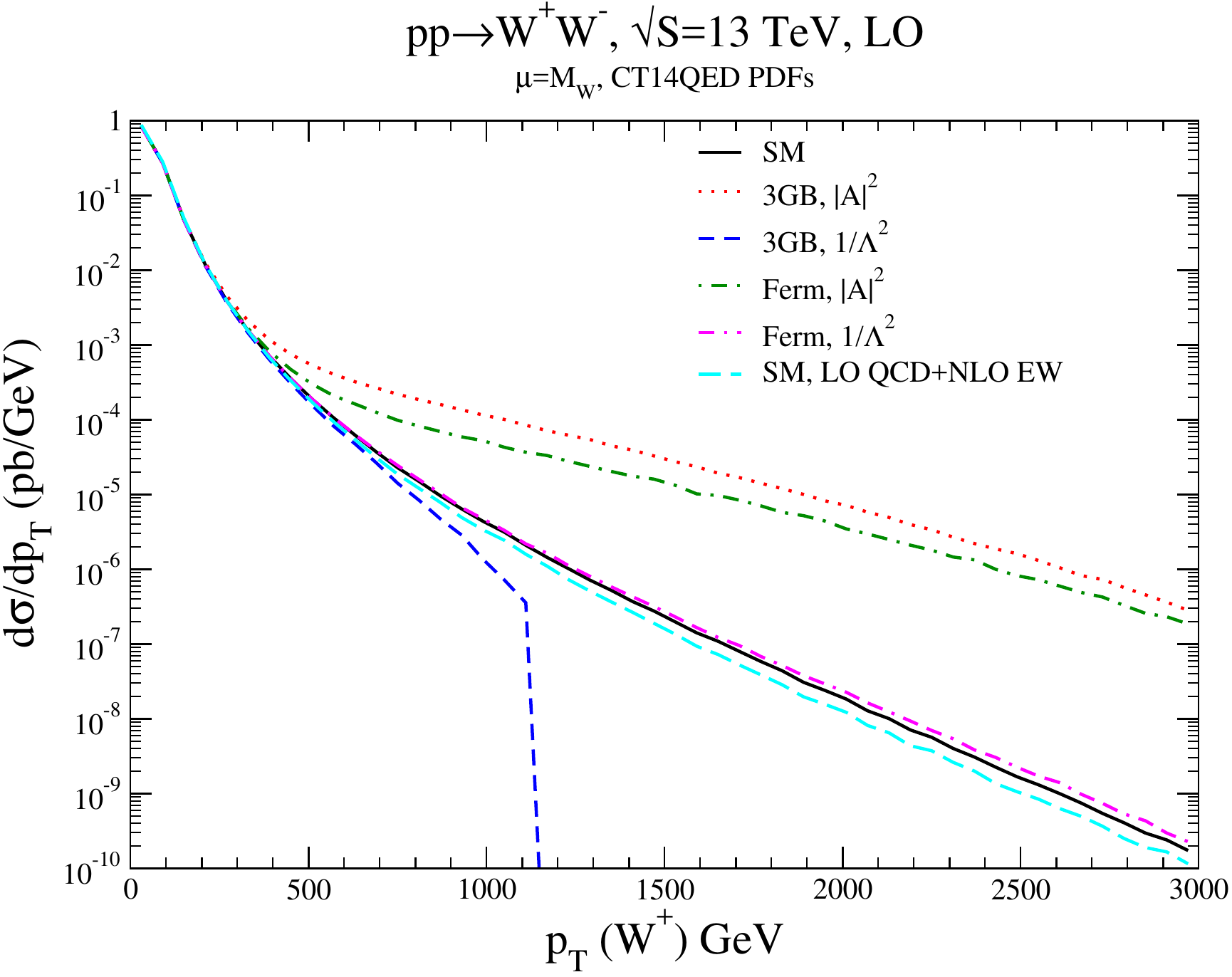}}
 \caption{Tree-level cross sections for $W^+_LW^-_L$ (LHS) and the sum
   of all polarizations (RHS) at the 13 TeV LHC in the SM and in the
   scenarios of Eq.(\ref {eq:fitparm}). The curves labelled $|A|^2$
   include the square of the dimension-$6$ amplitudes, while the
   curves labelled $1/\Lambda^2$ have the EFT result consistently
   truncated at  $1/\Lambda^2$. The LL (LHS)  and total (RHS) curves
   for the 3GB amplitude in the EFT should be truncated at $p_T\sim
   350$~GeV, where the LL rate becomes negative and the EFT expansion
   fails. The SM curve on the RHS includes the complete set of
   electroweak corrections.}
   \label{fig:ptamps2S}
\end{figure}
In Figs.~\ref{fig:ptamps} and \ref{fig:ptamps2S}, we show the
tree-level cross sections for transverse-transverse (TT),
transverse-longitudinal (TL+LT or simply LT), and
longitudinal-longitudinal (LL) $W^+W^-$ polarizations, along with the
sum over polarizations. We use CT14QED$\_$inc
PDFs~\cite{Schmidt:2015zda} implemented via
LHAPDF~\cite{Buckley:2014ana}, and set the renormalization and
factorization scales to be $M_W$. For the anomalous coupling
scenarios, we present both the amplitude-squared using the amplitudes
given in Appendix~\ref{appendixA}, and the EFT result consistently
truncated at ${\cal{O}}(1/\Lambda^2)$. The TT amplitude is by far the
largest contribution to the SM rate. As shown in the left-hand side
(LHS) of Fig.~\ref{fig:ptamps}, the $W^+_TW^-_T$ rates in our
anomalous coupling scenarios are indistinguishable from the SM when
truncating at ${\cal{O}}(1/\Lambda^2)$; the complete effect of the 3GB
anomalous couplings (red line on the LHS of Fig.~\ref{fig:ptamps})
comes from the square of the anomalous coupling contribution. The
growth of the TT amplitude with energy is due to the non-zero
$\lambda_{}^\gamma=\lambda_{}^Z ~(C_{3W})$. The right-hand side (RHS)
of Fig.~\ref{fig:ptamps} has the contribution from $W^\pm_L W^\mp_T$
production. The ``3GB'' scenario shows the EFT ${\cal{O}}(1/\Lambda^2)$
Born contributions becoming negative at $p_T\sim 500$~GeV, indicating
the failure of the EFT dimension-$6$ approximation for these
parameters. A comparison of  the blue and red lines on the RHS of
Fig.~\ref{fig:ptamps} illustrates the huge numerical impact of
including the full amplitude-squared, as compared to the $1/\Lambda^2$
truncation.  Similarly, at $1/\Lambda^2$ the anomalous fermion
coupling contribution is indistinguishable from the SM and their full
effect occurs at the amplitude-squared level.

The LL contribution is shown on
the LHS of Fig.~\ref{fig:ptamps2S} and at high $p_T$, we see the
growth of the amplitude-squared in both the ``3GB'' and ``Ferm'' scenarios.
The effects from anomalous gauge boson couplings and from anomalous
fermion couplings  are numerically very similar in the scenarios we
have chosen here, and the effects cannot be separated by a measurement
of $W^+W^-$ production alone. As in the LT case, we see that
truncating the 3GB rate at ${\cal{O}}(1/\Lambda^2)$ leads to negative
cross sections at small $p_T$.

The unpolarized cross sections are shown on the RHS of
Fig.~\ref{fig:ptamps2S}. The green and red curves on
the RHS of Fig.~\ref{fig:ptamps2S} demonstrate that the growth of the
cross section at high $p_T$ results from the square of the anomalous
coupling contribution. This contribution is formally of dimension-$8$
and is potentially of the same size as the neglected dimension-$8$
contributions.  The  RHS of Fig.~\ref{fig:ptamps2S} also shows the
effect of adding the complete electroweak corrections to the SM
prediction~\cite{Baglio:2013toa}.  Even at $p_T\sim 3$~TeV, these
corrections are small and so are neglected in the rest of this work.

\begin{figure}
  \centering
\subfloat{\includegraphics[width=0.48\textwidth]{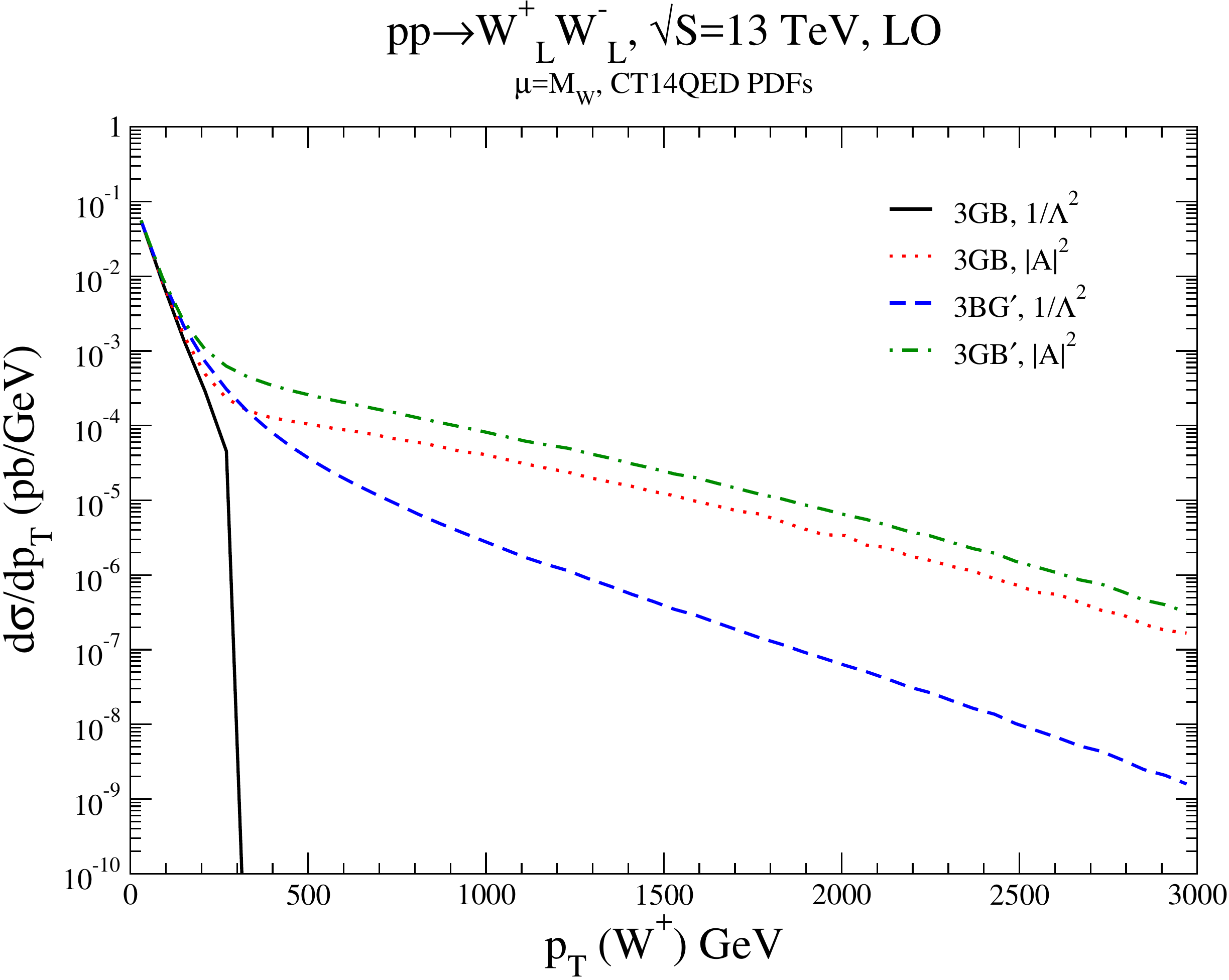}}
\subfloat{\includegraphics[width=0.48\textwidth]{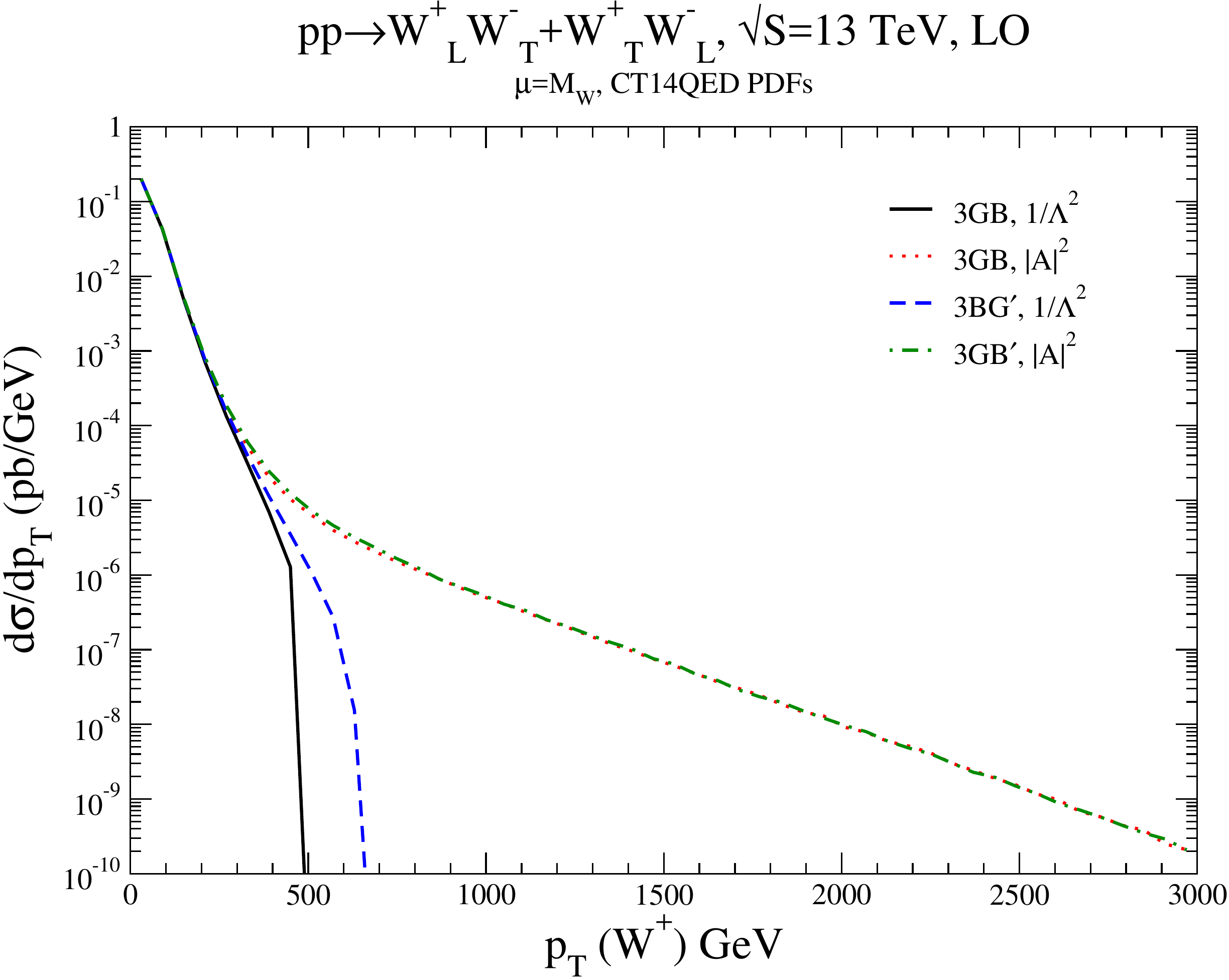}}
 \caption{Tree-level cross sections for $W^+_LW^-_L$ (LHS) and
   $W^+_TW^-_L+W^+_LW^-_T$  (RHS) at the 13 TeV LHC in the ``3GB'' and
   ``3GB$'$'' scenarios of Eqs.(\ref
   {eq:fitparm},\ref{eq:fitparm2}). The curves labelled $|A|^2$
   include the square of the dimension-$6$ amplitudes, while the
   curves labelled $1/\Lambda^2$ have the EFT result consistently
   truncated at  $1/\Lambda^2$. For the ``3GB'' amplitude, the LL
   (LHS) curves should be truncated at $p_T\sim350$~GeV and the LT
   (RHS) at $p_T\sim 500$~GeV, where the respective rate becomes
   negative and the EFT expansion fails. For the ``3GB$'$'' amplitude,
   the LT(RHS) curves are negative and should be truncated at $p_T\sim
   650$~GeV, while the LL (LHS) rates do not go negative.}
   \label{fig:ptampscomp}
\end{figure}

As the previous discussion indicates, for the parameter point ``3GB''
the EFT approximation begins to fail at a $W^+$ transverse momentum of
a few 100 GeV. However, where the EFT fails strongly depends the 
values of the anomalous triple gauge boson couplings. We consider
another scenario:
\begin{eqnarray}
  {\rm 3GB}':&\qquad \qquad &\delta g_1^Z =0.00452,\,
                             \delta\kappa_{}^Z=0.0239,\,
                             \lambda_{}^Z=0.0163,\nonumber
\label{eq:fitparm2}
\end{eqnarray}
where the anomalous fermionic couplings are set to
zero. Figure~\ref{fig:ptampscomp} compares the ``3GB'' and ``3GB$'$''
scenarios at leading order. The LL production rate (LHS) in the
``3GB$'$'' scenario does not go negative and the EFT approximation is
valid. This is to be compared to the ``3GB'' scenario where the EFT
approximation fails at $p_T\sim 350$~GeV. For the LT case, the
``3GB$'$'' rate becomes negative at $p_T\sim650$~GeV.  This extends
the validity of the EFT by $\sim150$~GeV above where the ``3GB''
scenario fails.
\subsection{NLO QCD Effects} 
\label{sec:nlo}
The lowest order results can potentially be significantly changed by
the inclusion of higher order QCD and EW effects. The EFT
contributions parametrized in the Lagrangians of
Eqs.~(\ref{eq:lagdef}) and (\ref{eq:dgdef}) do not affect the
structure of the QCD corrections. We can therefore include the NLO QCD
effects to ${\cal O}(\alpha_s^{})$ by calculating the virtual and real
contributions using the SM Lagrangian supplemented by  the anomalous
coupling terms, using the same structure for the Catani-Seymour
dipoles~\cite{Catani:1996vz} to cancel the infrared divergences as in
the SM calculation. Hence we use the same setup as in
Ref.~\cite{Baglio:2013toa} in which the details for the SM calculation
are given. The amplitudes for the NLO QCD EFT contributions have been
calculated using {\tt FeynArts-3.7}~\cite{Hahn:2000kx} and {\tt
  FormCalc-7.5}~\cite{Hahn:1998yk}, based on our Model File for {\tt
  FeynArts} for the anomalous couplings developed with the help of
{\tt FeynRules}~\cite{Alloul:2013bka}\footnote{
The open-access version of our code as well as the code giving the EW
corrections in the SM and developed in Ref.~\cite{Baglio:2013toa} are
included in
\url{https://quark.phy.bnl.gov/Digital_Data_Archive/dawson/ww_17}.}.
The one-loop integrals have been implemented with {\tt
  LoopTools-2.12}~\cite{Hahn:1998yk,vanOldenborgh:1990yc} and the Born
and virtual pieces have been cross-checked against an independent
analytical calculation. Since the SM NLO EW corrections are small (see
RHS of  Fig.~\ref{fig:ptamps2S}), we do not anticipate that the ${\cal
  O}(\delta g_{\rm EFT}^{}\alpha)$ corrections\footnote{$\delta g_{\rm
    EFT}^{}$ is here generically the deviation of any coupling from
  its SM value.} will be significant enough to deserve further
scrutiny.

\begin{figure}
  \centering
\subfloat{\includegraphics[width=0.48\textwidth]{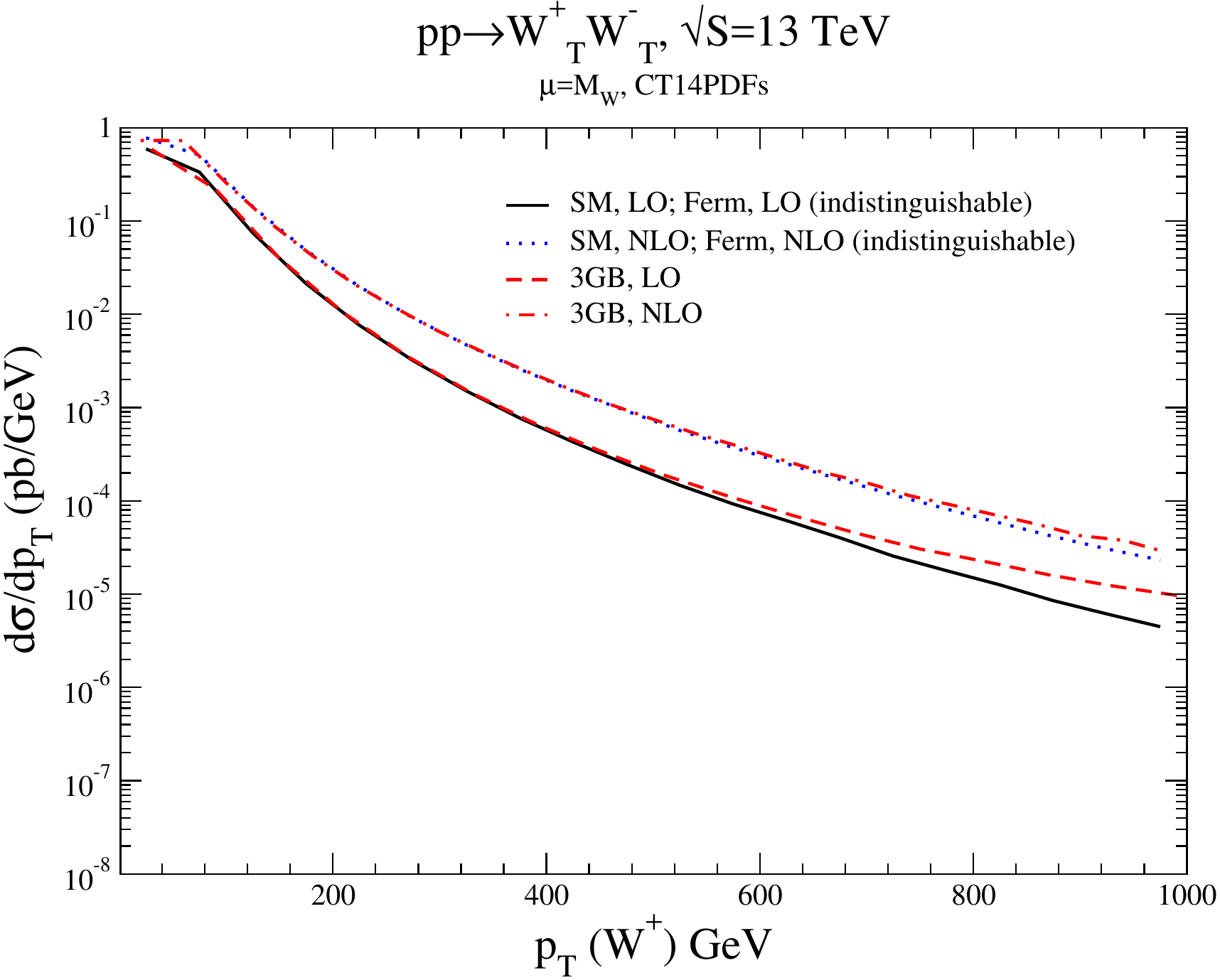}}
\subfloat{\includegraphics[width=0.48\textwidth]{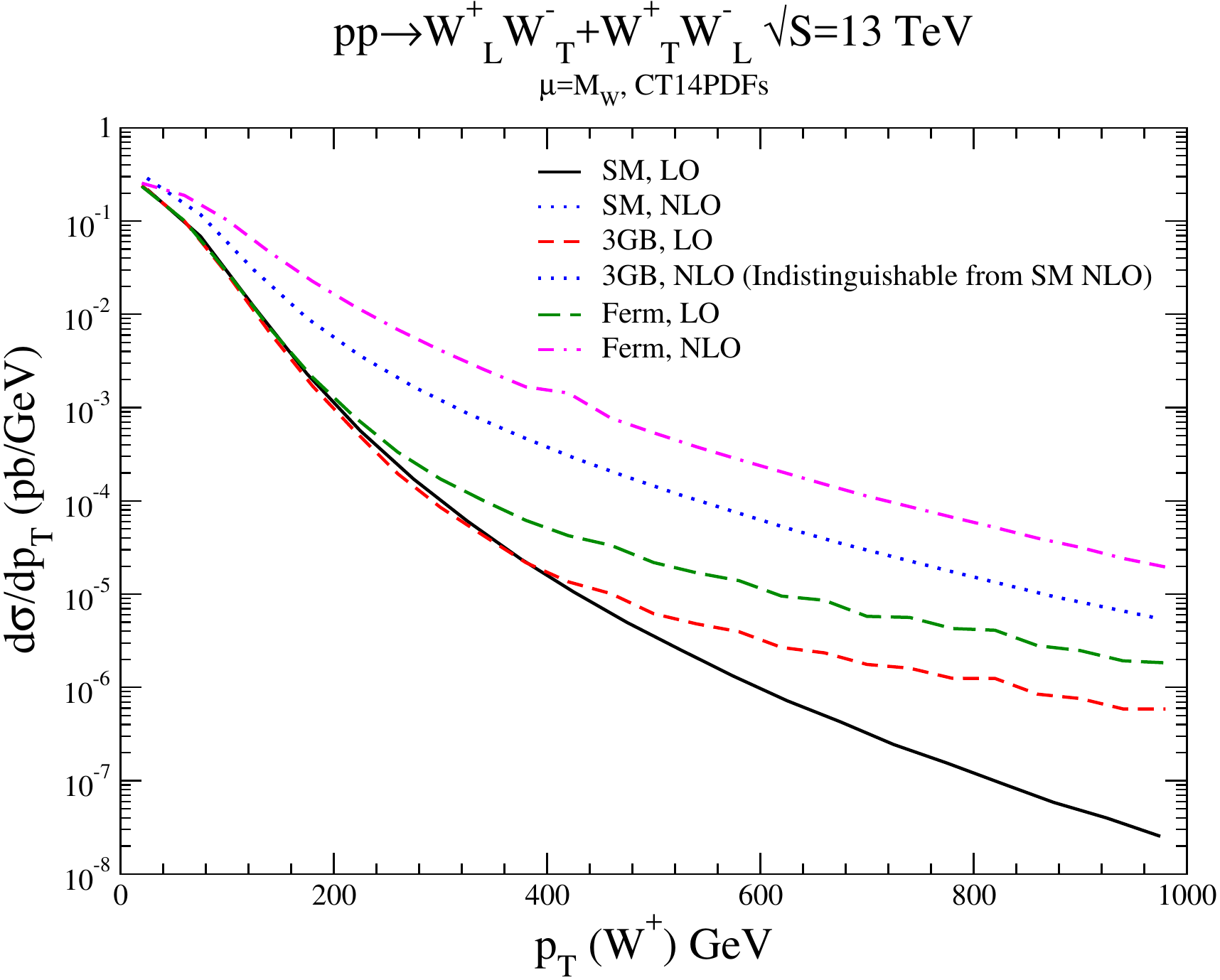}}
 \caption{Comparison of LO and NLO QCD results for the SM and the
   ``3GB'' and ``Ferm'' scenarios defined in Eqs. (\ref{eq:fitparm}),
   (\ref{eq:eftparms3GB}), and (\ref{eq:eftparmsFerm}) for $W^+_TW^-_T$
   (LHS) and $W^\pm_L W^\mp_T$ (RHS) productions. This figure includes
   the complete amplitude-squared.}
   \label{fig:ptamps1}
\end{figure}

\begin{figure}
  \centering
\subfloat{\includegraphics[width=0.48\textwidth]{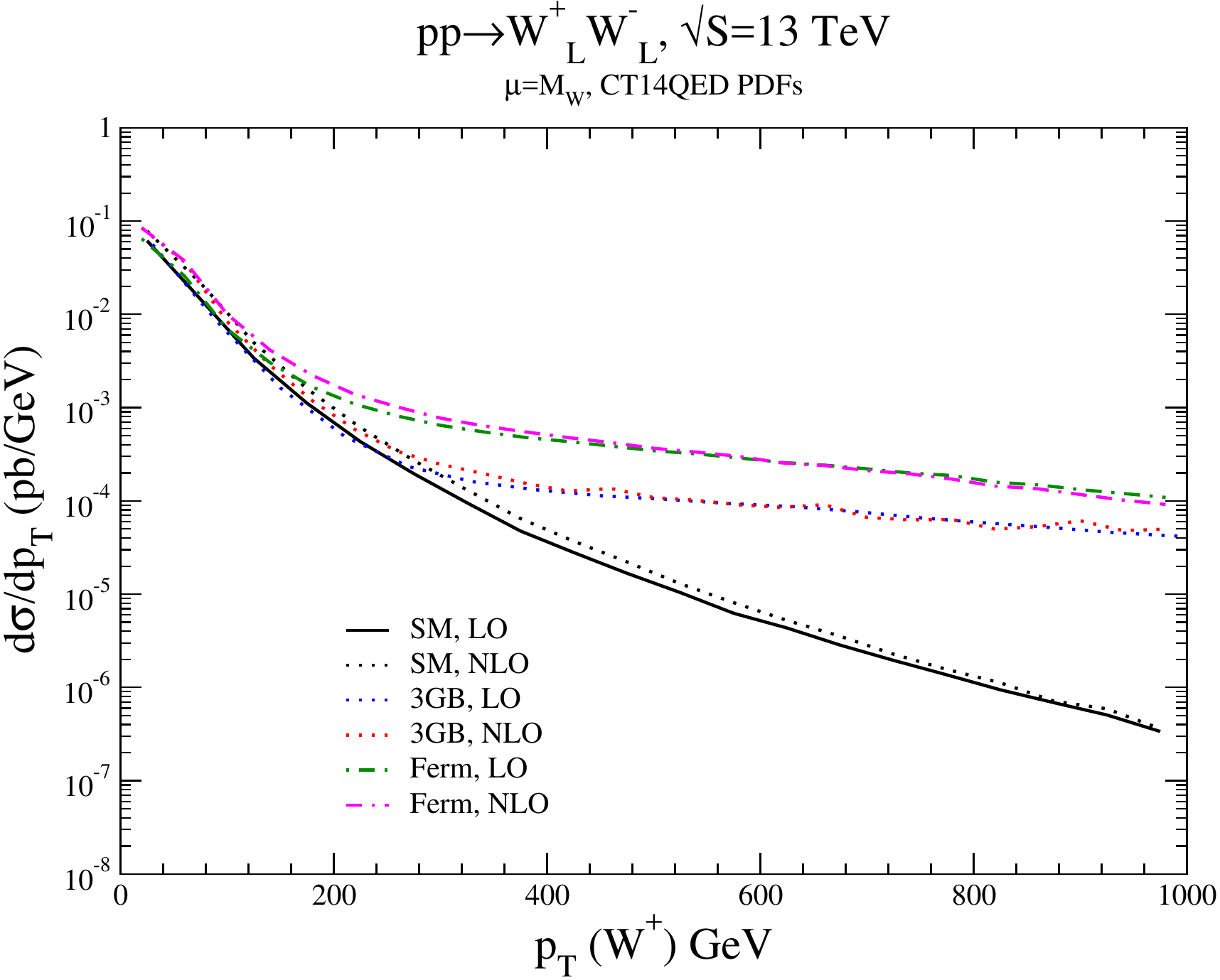}}
\subfloat{\includegraphics[width=0.48\textwidth]{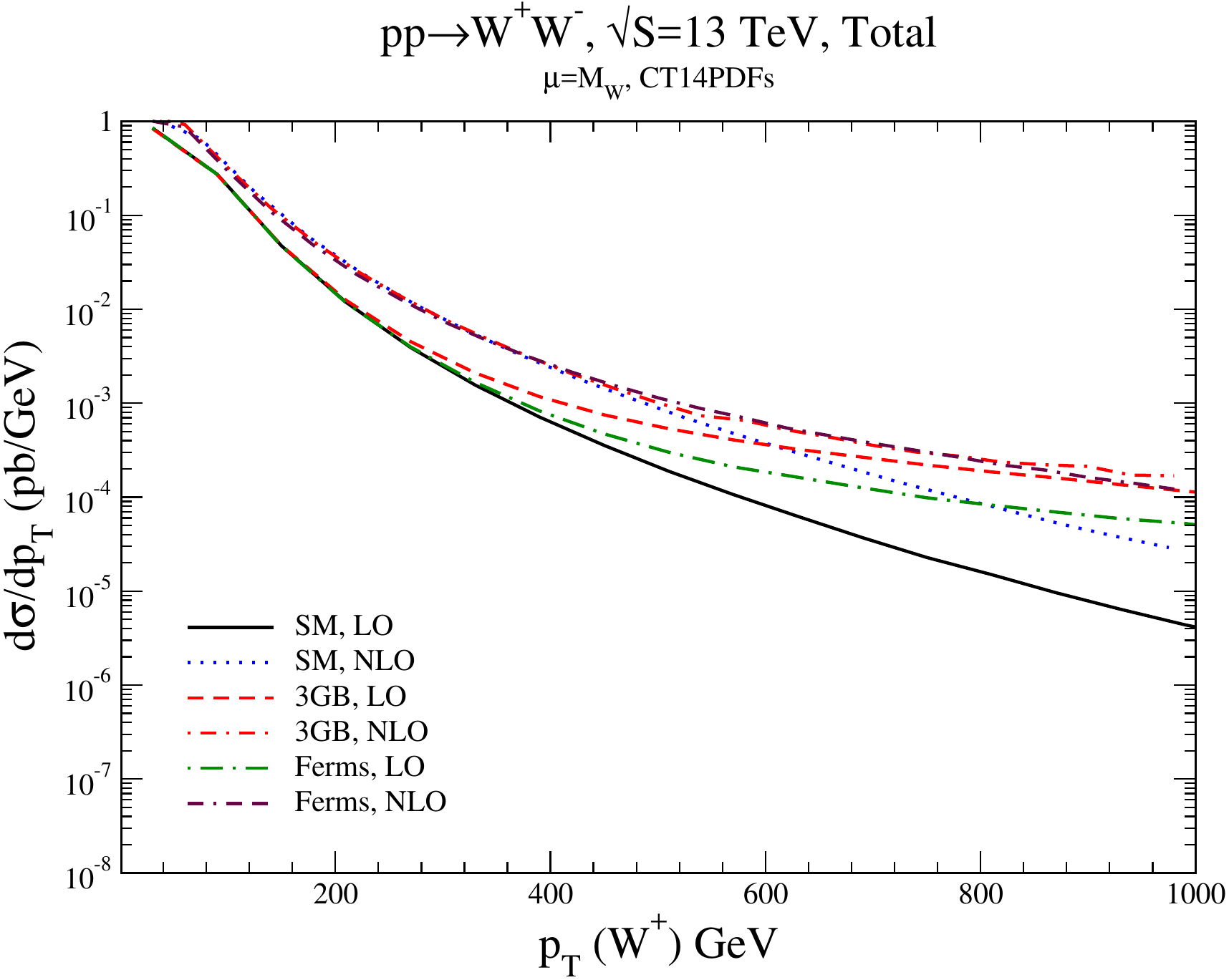}}
 \caption{Comparison of LO and NLO QCD results for the SM and the
   ``3GB'' and ``Ferm'' scenarios defined in Eqs. (\ref{eq:fitparm}),
   (\ref{eq:eftparms3GB}), and (\ref{eq:eftparmsFerm}) for
   $W^+_LW^-_L$ (LHS) and unpolarized $W^+_{} W^-_{}$ (RHS)
   productions. This figure includes the complete amplitude-squared.}
 \label{fig:ptamps2}
\end{figure}

The LO results presented in the previous section have emphasized that
the consistent EFT expansion up to ${\cal O}(1/\Lambda^2)$ can give
sizable deviations from the SM distributions, especially for the 3GB
operators (see Fig.~\ref{fig:ptamps2S}), and that the EFT expansion
truncated at  ${\cal O}(1/\Lambda^2)$  typically fails at moderate
$p_T$. The large effects from anomalous couplings result from the
terms which are quadratic in the squared-amplitudes. This observation
is  not significantly altered by the inclusion of the NLO QCD corrections. 

Figs.~\ref{fig:ptamps1} and \ref{fig:ptamps2}  show the LO and NLO QCD
corrected results for the TT, TL+LT, LL, and total $W^+W^-$ $p_T$
spectrums for the ``3GB'' and ``Ferm'' scenarios of the previous section
compared with the SM, when the total amplitudes are squared and all
terms included. The TT cross section has significant K factors for the
3GB anomalous couplings, the ``Ferm'' scenario, and the SM.  At NLO
the SM and ``Ferm'' scenarios are indistinguishable and most of the
3GB excess is erased (LHS of Fig.~\ref{fig:ptamps1}). The LT
polarization (RHS of Fig. \ref{fig:ptamps1}) displays a large
$K$-factor in the SM, ``3GB'', and ``Ferm'' scenarios, larger than
that for either the TT or LL polarizations. This behavior is due to
the fact that the SM amplitude-squared is suppressed by ${M_W^2\over
  s}$, enhancing naturally the $K$-factor. At NLO in the LT channel,
the enhancement of the rate in the ``Ferm'' scenario over the SM at
high $p_T$ persists; however, the ``3GB'' spectrum is
indistinguishable from the SM. Including the QCD corrections in the LT
channel is clearly critical for obtaining accurate
results. Interestingly, in the LL channel, the NLO QCD corrections in
all scenarios are small and the excesses remain. Finally, we consider
the total rate (RHS of Fig. \ref{fig:ptamps1}). The SM
  NLO QCD corrections are dominated by the TT channel, as could be
  espected by the large Sudakov logarithms coming from a hard $p_T^{}$
  jet radiating off a soft $W$ boson; the quarks being massless, the
  longitudinal $W$ bosons decouple at high energy.
At NLO, the effects
of the ``Ferm'' and ``3GB'' scenarios are largely similar, and the
enhancement relative to the SM stays intact.  The ``Ferm'' scenario
and SM have significant K-factors, while the NLO corrections to 3GB are
unimportant at high $p_T$. Since the LL configuration dominates the
``3GB'' scenario at high $p_T$, this conclusion was expected already
from the analysis of the LL curves. It should be noted that the
effects of the fermion anomalous couplings are largest in the LL
polarization. Hence, sensitivity to these couplings would be greatly
enhanced by performing an LL polarized analysis.

\begin{figure}
  \centering
  \subfloat{\includegraphics[width=0.48\textwidth]{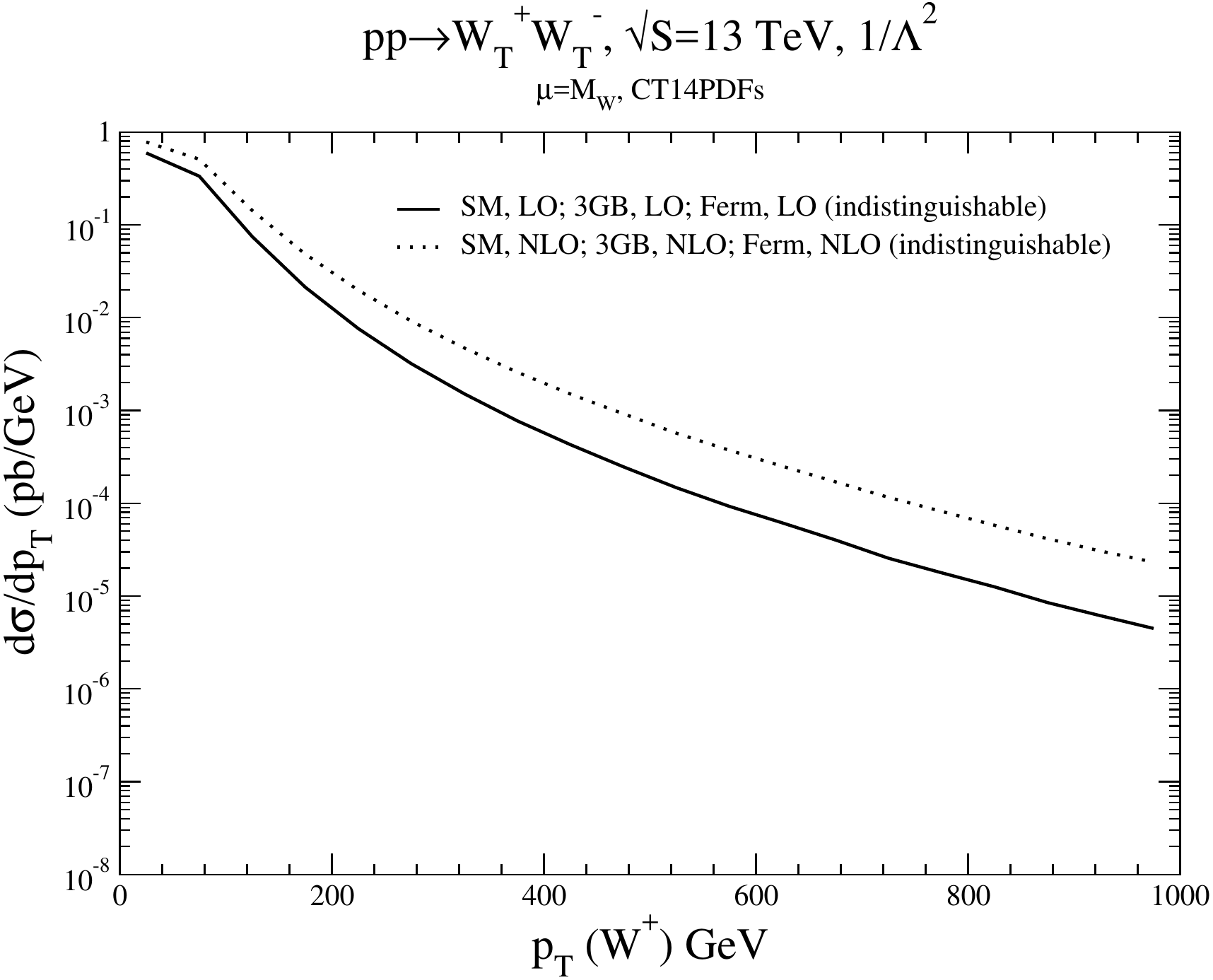}}
  \subfloat{\includegraphics[width=0.48\textwidth]{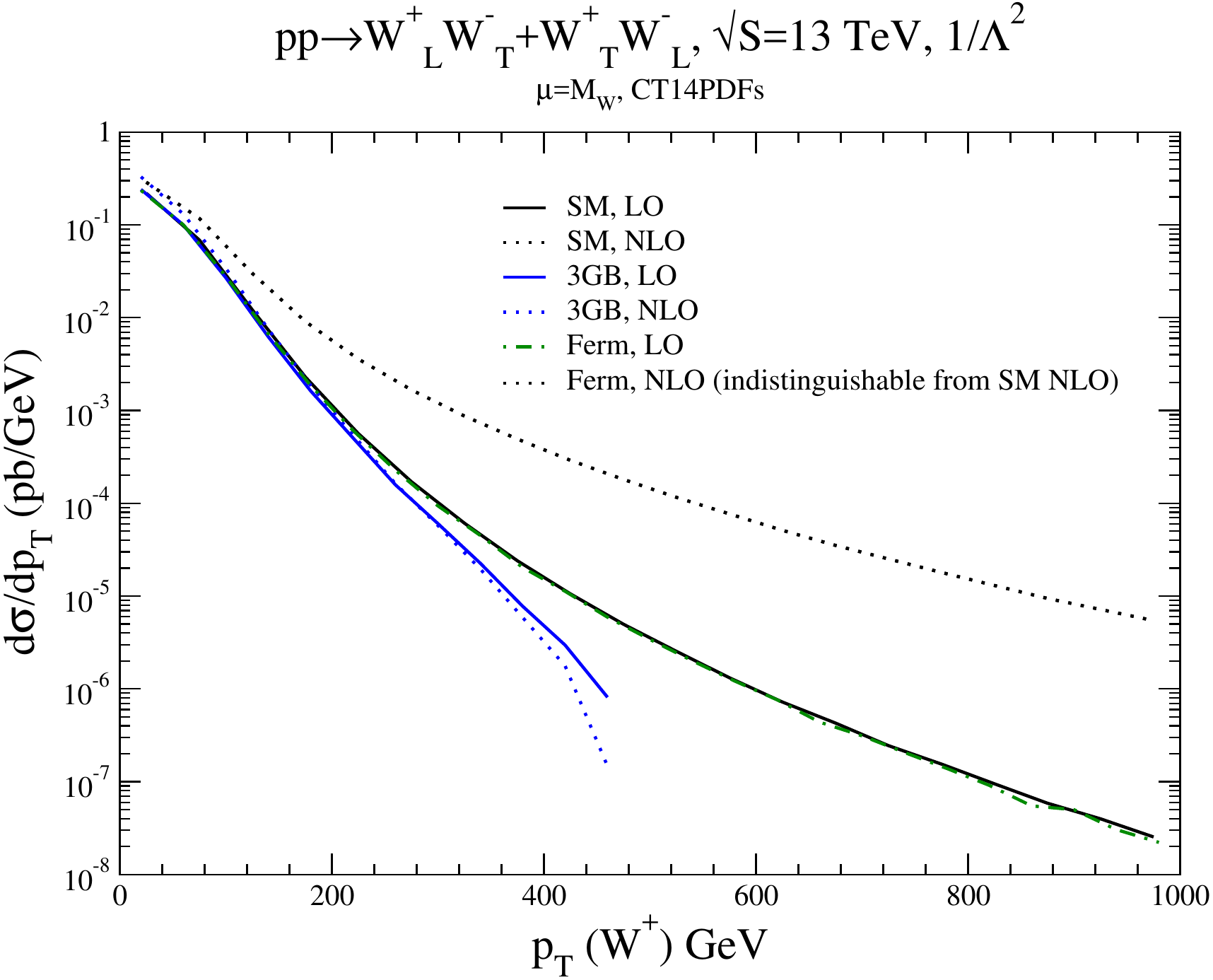}}
  \caption{Comparison of LO and NLO QCD results (LHS: TT polarization;
    RHS: LT+TL polarization) for the SM and the ``3GB'' and ``Ferm''
    scenarios defined in Eqs.(\ref{eq:fitparm}),
    (\ref{eq:eftparms3GB}), and (\ref{eq:eftparmsFerm}), truncated at
    ${\cal O}(1/\Lambda^2)$. The 3GB LT curve is truncated at $p_T\sim
    500$~GeV, since the LO cross section becomes negative at this
    point, signaling a breakdown in the EFT ${\cal O}(1/\Lambda^2)$
    approximation.}
  \label{fig:ptamps1_eft}
\end{figure}

\begin{figure}
  \centering
  \subfloat{\includegraphics[width=0.48\textwidth]{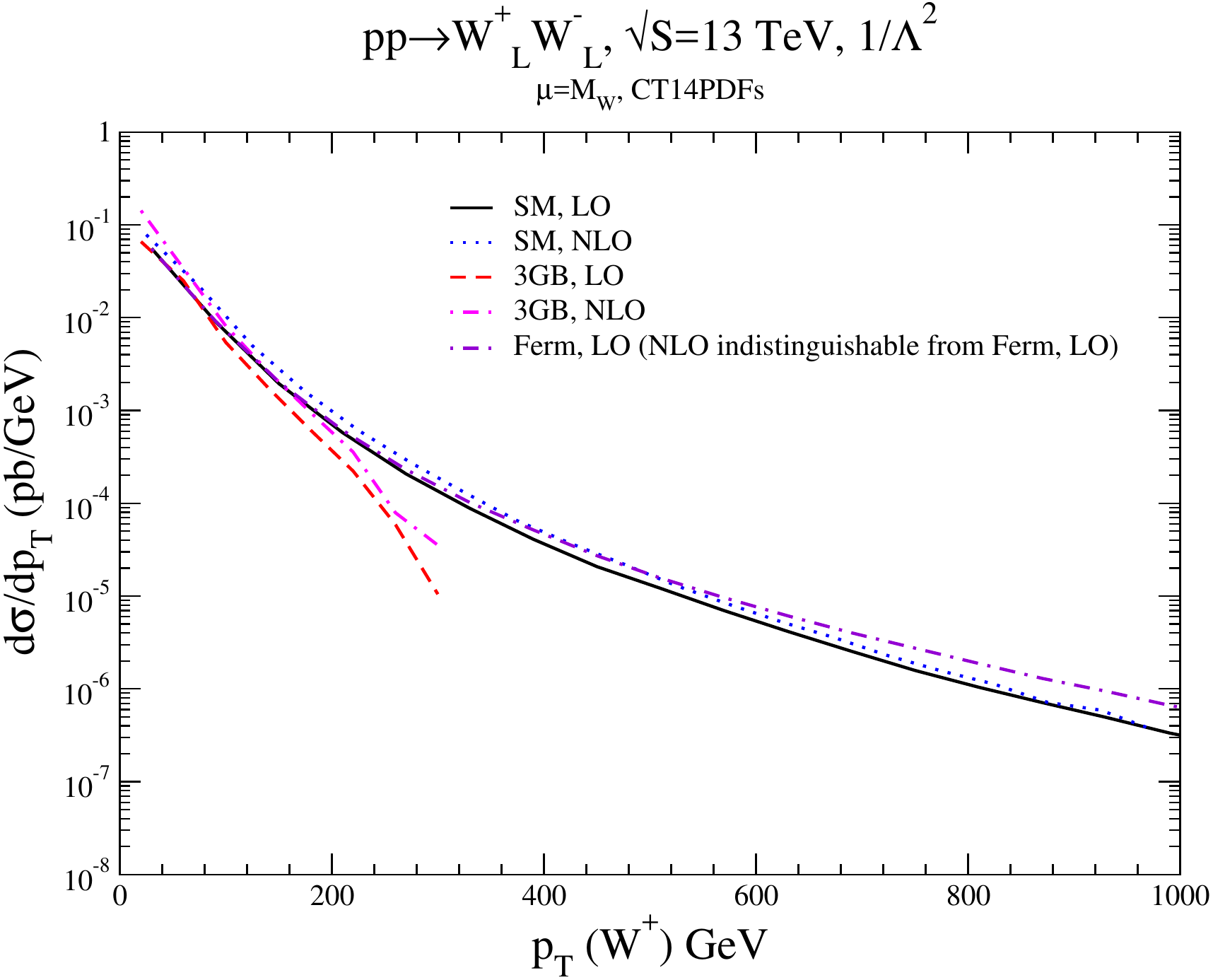}}
  \subfloat{\includegraphics[width=0.48\textwidth]{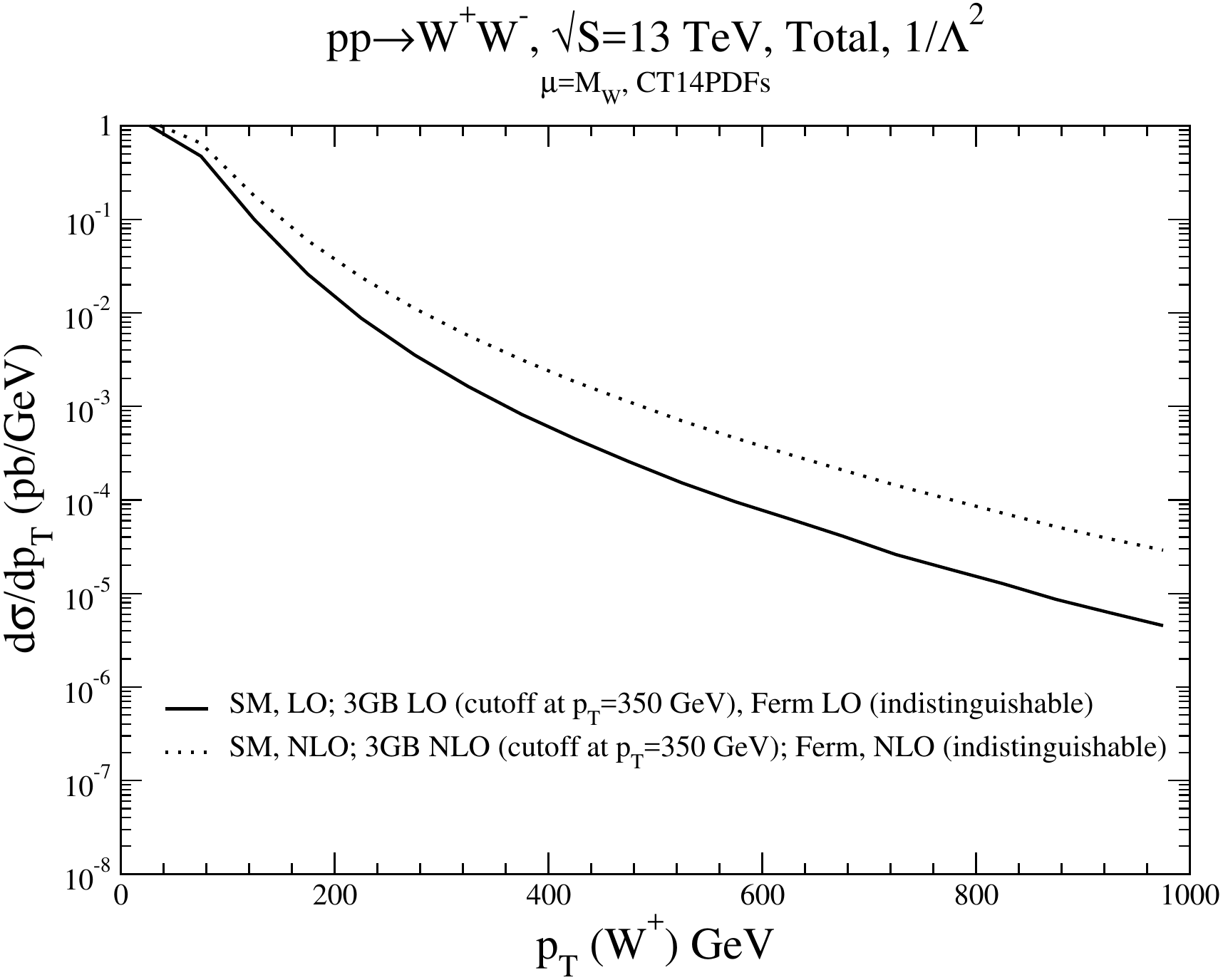}}
  \caption{Comparison of LO and NLO QCD results (LHS: LL polarization;
    RHS: unpolarized results) for the SM and the ``3GB'' and ``Ferm''
    scenarios defined in Eqs.(\ref{eq:fitparm}),
    (\ref{eq:eftparms3GB}), and (\ref{eq:eftparmsFerm}), truncated at
    ${\cal O}(1/\Lambda^2)$. The 3GB curves are truncated at $p_T\sim
    350$~GeV, since the LO LL cross section becomes negative at this
    point, signaling a breakdown in the EFT ${\cal O}(1/\Lambda^2)$
    approximation.}
  \label{fig:ptamps2_eft}
\end{figure}

In Figs.~\ref{fig:ptamps1_eft} and \ref{fig:ptamps2_eft}, we show the
comparison of the SM, ``3GB'' and ``Ferm'' scenarios at LO and NLO QCD,
truncated at ${\cal O}(1/\Lambda^2)$.  We have cut off the curves at
the points where the LO rates go negative for each polarization in
this approximation, since the EFT is no longer valid. It is
immediately apparent that the effects of the anomalous couplings are
small in the TT and LT polarizations in this EFT approximation and
that the entire effect in Fig.~\ref{fig:ptamps1} is from the
contributions quadratic in the anomalous couplings. In the LL
polarization (LHS of Fig.~\ref{fig:ptamps2_eft}), we see that as
before, the large enhancements seen earlier also arise from terms
quadratic in the anomalous couplings, although the small LO effect of
the anomalous fermion couplings at high $p_T$ does remain at NLO. The
RHS of Fig.~\ref{fig:ptamps2_eft} reflects the dominance of the TT
polarization and illustrates the necessity of extracting polarized
contributions. It also shows that the breakdown of the EFT ${\cal
  O}(1/\Lambda^2)$ expansion happens much earlier than would be
expected by the global fit analysis using 
$\sigma^{cut}$ of Eq. (\ref{eq:cut}).

\section{Conclusions} 
\label{sec:conc}
We have considered the effects of including both anomalous fermion and anomalous gauge boson interactions to 
the $p_T$ spectrum of $W^+W^-$ pair production. At LO QCD, the inclusion of even small anomalous fermion couplings
can significantly affect the fits extracted by ATLAS and CMS.  This observation has supported the need for global fits to
the spectrum of EFT couplings.  When the NLO QCD corrections are included, however, the sensitivity to anomalous fermion couplings
in  $W^+W^-$ pair production diminishes; although, a polarized analysis in the longitudinal-longitudinal mode could enhance sensitivity to these couplings.
The effect of the NLO QCD corrections is largest in the channel with one longitudinal and one transverse gauge boson, since the SM rate is 
suppressed by $M_W^2/s$ in the high energy limit. The NLO EW corrections are small and do not affect the fits to the anomalous couplings.
We have also re-iterated the well known observation that the sensitivity to anomalous couplings in $W^+W^-$ pair production results almost entirely 
from contributions quadratic in the dimension-$6$ EFT couplings. This is not altered by the inclusion of NLO QCD corrections.

The decays of the $W$ bosons to fermion pairs are not included in our analysis.  It should be noted that when the decays of the W bosons are considered, different W helicities can interfere.  It was shown in Ref. \cite{Duncan:1985vj} 
that the $W$ helicities
can be extracted  in hadronic collisions by measuring the azimuthal angle between the plane of the $W$ decay products and the plane of
the incoming protons and virtual $W$'s.  This idea has been used recently to develop observables that are sensitive
 to the interference between the SM and BSM  helicity amplitudes in $WZ$ and $W\gamma$ production~\cite{Azatov:2017kzw,Panico:2017frx}.    This is the subject of on-going study for the $W^+W^-$ case.

\begin{acknowledgments}
SD is supported by the United States Department of Energy under Grant
Contract DE-SC0012704 and is grateful to the University of T\"ubingen,
where this work was started. IML was supported in part by the
University of Kansas General Research Fund allocation 2302091. J.B. is
supported by the Kepler Center of the University of T\"ubingen and in
part by the German Research Foundation (DFG) through the grant JA
1954/1 . Parts of this work were performed thanks to the support of
the State of Baden-W\"urttemberg through bwHPC and the DFG through the
grant no. INST 39/963-1 FUGG. Digital data related to our results can
be found
at~\url{https://quark.phy.bnl.gov/Digital_Data_Archive/dawson/ww_17}.
\end{acknowledgments}

\appendix
\section{Helicity Amplitudes for \boldmath $q {\overline{q}} \rightarrow W^+W^-$\label{appendixA}}
From Ref.~\cite{Hagiwara:1986vm}, for the process $\bar{q}_s
q_{s'}\rightarrow W^+_{\lambda} W^-_{\lambda'}$ , the leading order
helicity amplitudes are,
\begin{eqnarray}
\mathcal{A}_{ss'\lambda\lambda'}&=&\sqrt{2}\widetilde{\mathcal{A}}_{ss'\lambda\lambda'}\widetilde{d}_{ss'\lambda\lambda'}(-1)^{\Delta\lambda},
\end{eqnarray}
where $\Delta s = (s-s')/2$, $\Delta\lambda = (\lambda-\lambda')$,
$J={\rm max}(|\Delta s|,|\Delta\lambda|)$.  We can further decompose
the amplitudes, separating out the information from the quark
couplings,
\begin{eqnarray}
\widetilde{\mathcal{A}}_{-+\lambda\lambda'}&=&g_Z^2c_W^2\left(g^{Zq}_R +\delta g^{Zq}_R\right)\beta_W\frac{s}{s-M_Z^2}A^Z_{\lambda\lambda'}+e^2 Q_q \beta_W A_{\lambda\lambda'}^\gamma\nonumber\\
&=&e^2 Q_q \beta_W\left(A_{\lambda\lambda'}^\gamma-\frac{s}{s-M_Z^2}A_{\lambda\lambda'}^Z\right)+g^2 \delta g_R^{Zq}\beta_W\frac{s}{s-M_Z^2}A_{\lambda\lambda'}^Z\label{eq:HelRH}\\
\widetilde{\mathcal{A}}_{+-\lambda\lambda'}&=&g_Z^2c_W^2\left(g^{Zq}_L+\delta g^{Zq}_L\right)\beta_W\frac{s}{s-M_Z^2}A^{Z}_{\lambda\lambda'}+e^2 Q_q\beta_W A_{\lambda\lambda'}^\gamma+2\,T_3^q\frac{g^2}{\beta_W}\left(1+\delta g_L^W\right)^2A_{\lambda\lambda'}^W,\nonumber
\end{eqnarray}
with $\beta_W^{} = \sqrt{1-4 M_W^2 /s}$.

The $A^Z, A^\gamma$ and $A^W$ coefficients are ($V=\gamma, Z$ and
$\delta g_1^\gamma = 0$):
\begin{eqnarray}
A_{00}^V&=&\frac{s}{2M_W^2}+1+\left(\delta g_1^V+\delta\kappa^V\frac{s}{2M_W^2}\right),\nonumber\\
A_{00}^W&=&-\frac{s}{4M_W^2}+\frac{4M_W^2}{s}\frac{1}{1+\beta_W^2-4\,T_3^q\,\beta_W\cos\theta},\nonumber\\
A_{+0}^V&=&A_{0+}^V=A_{-0}^V=A_{0-}^V=\frac{\sqrt{s}}{M_W}\left(1+\frac{1}{2}\left(\delta g_1^V+\delta\kappa^V+
\lambda^V\right)\right),\nonumber\\
A_{+0}^W&=&A_{0-}^W=\frac{\sqrt{s}}{M_W}\left(\frac{2 M_W^2}{s}\frac{1}{1+\beta_W^2-4\,T_3^q\,\beta_W\cos\theta}(1-2\,T_3^q\,\beta_W)-\frac{1}{2}\right),\nonumber\\
A_{0+}^W&=&A_{-0}^W=\frac{\sqrt{s}}{M_W}\left(\frac{2 M_W^2}{s}\frac{1}{1+\beta_W^2-4\,T_3^q\,\beta_W\cos\theta}(1+2\,T_3^q\,\beta_W)-\frac{1}{2}\right),\nonumber\\
A_{--}^V&=&A_{++}^V=1+\delta g_1^V+\frac{s}{2M_W^2}\lambda^V,\nonumber\\
A_{--}^W&=&A_{++}^W=-\frac{1}{2}+\frac{2M_W^2}{s}\frac{1}{1+\beta_W^2-4\,T_3^q\,\beta_W\cos\theta},\nonumber\\
A_{+-}^V&=&A_{-+}^V=0,\nonumber\\
A_{+-}^W&=&A_{-+}^W=2\sqrt{2}\,T_3^q\,\beta_W\frac{1}{1+\beta_W^2-4\,T_3^q\,\beta_W\cos\theta}.
\end{eqnarray}

The necessary  Wigner-D functions are,   
\begin{eqnarray}
\widetilde{d}_{-1,+1,0,0}&=&\widetilde{d}_{-1,+1,-1,-1}=\widetilde{d}_{-1,+1,+1,+1}=-\widetilde{d}_{+1,-1,0,0}=-\widetilde{d}_{+1,-1,-1,-1}=-\widetilde{d}_{+1,-1,+1,+1}=\frac{1}{\sqrt{2}}\sin\theta,\nonumber\\
\widetilde{d}_{-1,+1,0,+1}&=&\widetilde{d}_{-1,+1,-1,0}=\widetilde{d}_{+1,-1,0,-1}=\widetilde{d}_{+1,-1,+1,0}=-\frac{1}{2}\left(1+\cos\theta\right),\nonumber\\
\widetilde{d}_{-1,+1,+1,0}&=&\widetilde{d}_{-1,+1,0,-1}=\widetilde{d}_{+1,-1,-1,0}=\widetilde{d}_{+1,-1,0,+1}=-\frac{1}{2}\left(1-\cos\theta\right),\nonumber\\
\widetilde{d}_{+1,-1,+1,-1}&=&-\widetilde{d}_{-1,+1,-1,+1}=\frac{1}{2}\left(1+\cos\theta\right)\sin\theta,\nonumber\\
\widetilde{d}_{-1,+1,+1,-1}&=&-\widetilde{d}_{+1,-1,-1,+1}=\frac{1}{2}\left(1-\cos\theta\right)\sin\theta.
\end{eqnarray}

\bibliographystyle{utphys}
\bibliography{ww}

\end{document}